\documentclass[twocolumn,numberedappendix,twocolappendix,appendixfloats]{openjournal_dhayaa}

\usepackage{amsmath}
\usepackage{fontawesome5}
\usepackage{color}
\usepackage{xcolor}
\usepackage{xspace}
\usepackage{enumitem}

\usepackage{natbib}
\setcitestyle{aysep={}}

\usepackage[colorlinks=true
  ,urlcolor=blue
  ,anchorcolor=blue
  ,citecolor=blue
  ,filecolor=blue
  ,linkcolor=blue
  ,menucolor=blue
  ,linktocpage=true
  ,pdfproducer=medialab
  ,pdfa=true
]{hyperref}

\usepackage{graphicx}	% Including figure files
\usepackage{amsmath}	% Advanced maths commands

\usepackage{fontawesome5}
\usepackage{color}
\usepackage{xcolor}

\newcommand{\eg}{{\sl e.g.}, }   
\newcommand{\ie}{{\sl i.e.}, }   
\newcommand{\etc}{{\sl etc.}\,}

\newcommand{\Om}{\Omega_{\rm m}}
\newcommand{\Seight}{S_8}
\newcommand{\LCDM}{$\Lambda$CDM\xspace}
\newcommand{\wCDM}{$w$CDM\xspace}

\newcommand{\decade}{DECADE\xspace}

\definecolor{orcidlogocol}{HTML}{A6CE39}
\definecolor{purple}{RGB}{128, 0, 128}
\definecolor{maroon}{RGB}{128, 0, 0}

\newcommand{\OrcidID}[1]{ \href[urlcolor = red]{https://orcid.org/#1}{\textcolor{lightgray}{\faOrcid}}}
\newcommand{\OrcidIDName}[2]{\href{https://orcid.org/#1}{#2}}

\defcitealias{y3-shapecatalog}{\textsc{GS21}}
\defcitealias{paper1}{\textsc{Paper I}}
\defcitealias{paper2}{\textsc{Paper II}}
\defcitealias{paper3}{\textsc{Paper III}}
\defcitealias{paper4}{\textsc{Paper IV}}

\newcommand*{\vcenteredhbox}[1]{\begingroup
\setbox0=\hbox{#1}\parbox{\wd0}{\box0}\endgroup}

\begin{document}
{\footnotesize \hfill FERMILAB-PUB-25-0066-LDRD-PPD}

% Title of the paper, and the short title which is used in the headers.
% Keep the title short and informative.
\title{The DECADE cosmic shear project IV: cosmological constraints from 107 million galaxies across 5,400 deg$^2$ of the sky}
\shortauthors{Anbajagane \& Chang et al.}
\shorttitle{The DECADE Cosmic Shear Project IV: Cosmological Constraints}

\author{\OrcidIDName{0000-0002-6021-8760}{D.~Anbajagane} (\vcenteredhbox{\includegraphics[height=1.2\fontcharht\font`\B]{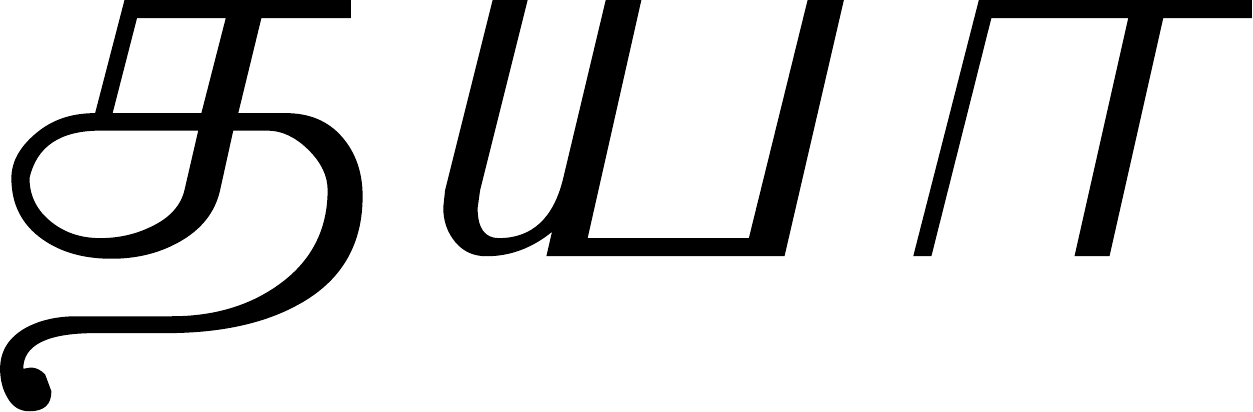}})$^\star$}
\affiliation{Department of Astronomy and Astrophysics, University of Chicago, Chicago, IL 60637, USA}
\affiliation{Kavli Institute for Cosmological Physics, University of Chicago, Chicago, IL 60637, USA}

\author{\OrcidIDName{0000-0002-7887-0896}{C.~Chang}$^\star$}
\affiliation{Department of Astronomy and Astrophysics, University of Chicago, Chicago, IL 60637, USA}
\affiliation{Kavli Institute for Cosmological Physics, University of Chicago, Chicago, IL 60637, USA}

\author{\OrcidIDName{0000-0001-8251-933X}{A.~Drlica-Wagner}}
\affiliation{Fermi National Accelerator Laboratory, P. O. Box 500, Batavia, IL 60510, USA}
\affiliation{Department of Astronomy and Astrophysics, University of Chicago, Chicago, IL 60637, USA}
\affiliation{Kavli Institute for Cosmological Physics, University of Chicago, Chicago, IL 60637, USA}

\author{\OrcidIDName{0000-0003-0478-0473}{C.~Y.~Tan}}
\affiliation{Department of Physics, University of Chicago, Chicago, IL 60637, USA}
\affiliation{Kavli Institute for Cosmological Physics, University of Chicago, Chicago, IL 60637, USA}

\author{\OrcidIDName{0000-0002-6904-359X}{M.~Adamow}}
\affiliation{Center for Astrophysical Surveys, National Center for Supercomputing Applications, 1205 West Clark St., Urbana, IL 61801, USA}
\affiliation{Department of Astronomy, University of Illinois at Urbana-Champaign, 1002 W. Green Street, Urbana, IL 61801, USA}

\author{\OrcidIDName{0000-0002-4588-6517}{R.~A.~Gruendl}}
\affiliation{Center for Astrophysical Surveys, National Center for Supercomputing Applications, 1205 West Clark St., Urbana, IL 61801, USA}
\affiliation{Department of Astronomy, University of Illinois at Urbana-Champaign, 1002 W. Green Street, Urbana, IL 61801, USA}

\author{\OrcidIDName{0000-0002-6002-4288}{L.~F.~Secco}}
\affiliation{Kavli Institute for Cosmological Physics, University of Chicago, Chicago, IL 60637, USA}

\author{\OrcidIDName{0000-0002-7523-582X}{Z.~Zhang}}
\affiliation{Department of Astronomy and Astrophysics, University of Chicago, Chicago, IL 60637, USA}
\affiliation{Department of Physics, Stanford University, 382 Via Pueblo Mall, Stanford, CA 94305, USA}
\affiliation{SLAC National Accelerator Laboratory, Menlo Park, CA 94025, USA}

\author{\OrcidIDName{0000-0001-7774-2246}{M.~R.~Becker}}
\affiliation{Argonne National Laboratory, 9700 South Cass Avenue, Lemont, IL 60439, USA}

\author{\OrcidIDName{0000-0001-6957-1627}{P.~S.~Ferguson}}
\affiliation{DIRAC Institute, Department of Astronomy, University of Washington, 3910 15th Ave NE, Seattle, WA, 98195, USA}

\author{\OrcidIDName{0009-0005-1143-495X}{N.~Chicoine}}
\affiliation{Department of Astronomy and Astrophysics, University of Chicago, Chicago, IL 60637, USA}
\affiliation{Department of Physics and Astronomy, University of Pittsburgh, 3941 O’Hara Street, Pittsburgh, PA 15260}

\author{\OrcidIDName{0000-0003-4394-7491}{K.~Herron}}
\affiliation{Department of Physics and Astronomy, Dartmouth College, Hanover, NH 03755, USA}

\author{\OrcidIDName{0000-0001-8505-1269}{A.~Alarcon}}
\affiliation{Institute of Space Sciences (ICE, CSIC),  Campus UAB, Carrer de Can Magrans, s/n,  08193 Barcelona, Spain}

\author{\OrcidIDName{0000-0002-5279-0230}{R.~Teixeira}}
\affiliation{Department of Astronomy and Astrophysics, University of Chicago, Chicago, IL 60637, USA}
\affiliation{Department of Physics, Duke University Durham, NC 27708, USA}

\author{\OrcidIDName{0000-0003-2911-2025}{D.~Suson}}
\affiliation{Department of Chemistry and Physics, Purdue University Northwest 2200, 169th Ave, Hammond, IN 46323}

\author{\OrcidIDName{0000-0002-3173-2592}{A.~N.~Alsina}}
\affiliation{Instituto de Física Gleb Wataghin, Universidade Estadual de Campinas, 13083-859, Campinas, SP, Brazil}

\author{\OrcidIDName{0000-0002-6445-0559}{A.~Amon}}
\affiliation{Department of Astrophysical Sciences, Princeton University, Peyton Hall, Princeton, NJ 08544, USA}

\author{\OrcidIDName{0000-0003-0171-6900}{F.~Andrade-Oliveira}}
\affiliation{Physik-Institut, University of Zurich, Winterthurerstrasse 190, CH-8057 Zurich, Switzerland}

\author{\OrcidIDName{0000-0002-4687-4657}{J.~Blazek}}
\affiliation{Department of Physics, Northeastern University, Boston, MA 02115, USA}

\author{\OrcidIDName{0000-0003-4383-2969}{C.~R.~Bom}}
\affiliation{Centro Brasileiro de Pesquisas F\'isicas, Rua Dr. Xavier Sigaud 150, 22290-180 Rio de Janeiro, RJ, Brazil}

\author{\OrcidIDName{0000-0001-5871-0951}{H.~Camacho}}
\affiliation{Physics Department, Brookhaven National Laboratory, Upton, NY 11973}

\author{\OrcidIDName{0000-0002-3690-105X}{J.~A.~Carballo-Bello}}
\affiliation{Instituto de Alta Investigaci\'on, Universidad de Tarapac\'a, Casilla 7D, Arica, Chile}

\author{\OrcidIDName{0000-0003-3044-5150}{A.~Carnero~Rosell}}
\affiliation{Universidad de La Laguna, Dpto. Astrofísica, E-38206 La Laguna, Tenerife, Spain}
\affiliation{Instituto de Astrofisica de Canarias, E-38205 La Laguna, Tenerife, Spain}
\affiliation{Laborat\'orio Interinstitucional de e-Astronomia - LIneA, Rua Gal. Jos\'e Cristino 77, Rio de Janeiro, RJ - 20921-400, Brazil}

\author{\OrcidIDName{0000-0003-2965-6786}{R.~Cawthon}}
\affiliation{Physics Department, William Jewell College, Liberty, MO, 64068}

\author{\OrcidIDName{0000-0003-1697-7062}{W.~Cerny}}
\affiliation{Department of Astronomy, Yale University, New Haven, CT 06520, USA}

\author{\OrcidIDName{0000-0002-5636-233X}{A.~Choi}}
\affiliation{NASA Goddard Space Flight Center, 8800 Greenbelt Rd, Greenbelt, MD 20771, USA}

\author{\OrcidIDName{0000-0003-1680-1884}{Y.~Choi}}
\affiliation{NSF National Optical-Infrared Astronomy Research Laboratory, 950 North Cherry Avenue, Tucson, AZ 85719, USA}

\author{\OrcidIDName{0000-0002-8446-3859}{S.~Dodelson}}
\affiliation{Kavli Institute for Cosmological Physics, University of Chicago, Chicago, IL 60637, USA}
\affiliation{Fermi National Accelerator Laboratory, P. O. Box 500, Batavia, IL 60510, USA}
\affiliation{Department of Astronomy and Astrophysics, University of Chicago, Chicago, IL 60637, USA}

\author{\OrcidIDName{0000-0003-4480-0096}{C.~Doux}}
\affiliation{Université Grenoble Alpes, CNRS, LPSC-IN2P3, 38000 Grenoble, France}

\author{\OrcidIDName{0000-0002-1407-4700}{K.~Eckert}}
\affiliation{Department of Physics and Astronomy, University of Pennsylvania, Philadelphia, PA 19104, USA}

\author{\OrcidIDName{0000-0001-5148-9203}{J.~Elvin-Poole}}
\affiliation{Department of Physics and Astronomy, University of Waterloo, 200 University Ave W, Waterloo, ON N2L 3G1, Canada}

\author{\OrcidIDName{0000-0003-4373-2386}{J.~Esteves}}
\affiliation{Department of Physics, Harvard University, MA 02138, USA}

\author{\OrcidIDName{0000-0001-6134-8797}{M.~Gatti}}
\affiliation{Kavli Institute for Cosmological Physics, University of Chicago, Chicago, IL 60637, USA}

\author{\OrcidIDName{0000-0002-3730-1750}{G.~Giannini}}
\affiliation{Kavli Institute for Cosmological Physics, University of Chicago, Chicago, IL 60637, USA}

\author{\OrcidIDName{0000-0003-3270-7644}{D.~Gruen}}
\affiliation{University Observatory, Faculty of Physics, Ludwig-Maximilians-Universität, Scheinerstr. 1, 81679 Munich, Germany}
\affiliation{Excellence Cluster ORIGINS, Boltzmannstr. 2, 85748 Garching, Germany}

\author{\OrcidIDName{0000-0001-9994-1115}{W.~G.~Hartley}}
\affiliation{Department of Astronomy, University of Geneva, ch. d’Ecogia 16, 1290 Versoix, Switzerland}

\author{\OrcidIDName{0000-0001-6718-2978}{K.~Herner}}
\affiliation{Fermi National Accelerator Laboratory, P. O. Box 500, Batavia, IL 60510, USA}

\author{\OrcidIDName{0000-0002-9378-3424}{E.~M.~Huff}}
\affiliation{Jet Propulsion Laboratory, California Institute of Technology, 4800 Oak Grove Dr., Pasadena, CA 91109, USA}

\author{\OrcidIDName{0000-0001-5160-4486}{D.~J.~James}}
\affiliation{Applied Materials Inc., 35 Dory Road, Gloucester, MA 01930}
\affiliation{ASTRAVEO LLC, PO Box 1668, Gloucester, MA 01931}

\author{\OrcidIDName{0000-0002-4179-5175}{M.~Jarvis}}
\affiliation{Department of Physics and Astronomy, University of Pennsylvania, Philadelphia, PA 19104, USA}

\author{\OrcidIDName{0000-0001-8356-2014}{E.~Krause}}
\affiliation{Department of Astronomy/Steward Observatory, University of Arizona, Tucson, AZ 85721 USA}

\author{\OrcidIDName{0000-0003-2511-0946}{N.~Kuropatkin}}
\affiliation{Fermi National Accelerator Laboratory, P. O. Box 500, Batavia, IL 60510, USA}

\author{\OrcidIDName{0000-0002-9144-7726}{C.~E.~Mart\'inez-V\'azquez}}
\affiliation{International Gemini Observatory/NSF NOIRLab, 670 N. A'ohoku Place, Hilo, Hawai'i, 96720, USA}

\author{\OrcidIDName{0000-0002-8093-7471}{P.~Massana}}
\affiliation{NSF's NOIRLab, Casilla 603, La Serena, Chile}

\author{\OrcidIDName{0000-0003-3519-4004}{S.~Mau}}
\affiliation{Department of Physics, Stanford University, 382 Via Pueblo Mall, Stanford, CA 94305, USA}
\affiliation{Kavli Institute for Particle Astrophysics \& Cosmology, P.O.\ Box 2450, Stanford University, Stanford, CA 94305, USA}

\author{\OrcidIDName{0000-0002-4475-3456}{J.~McCullough}}
\affiliation{Department of Astrophysical Sciences, Peyton Hall, Princeton University, Princeton, NJ USA 08544}

\author{\OrcidIDName{0000-0003-0105-9576}{G.~E.~Medina}}
\affiliation{Dunlap Institute for Astronomy \& Astrophysics, University of Toronto, 50 St George Street, Toronto, ON M5S 3H4, Canada}
\affiliation{David A. Dunlap Department of Astronomy \& Astrophysics, University of Toronto, 50 St George Street, Toronto ON M5S 3H4, Canada}

\author{\OrcidIDName{0000-0001-9649-4815}{B.~Mutlu-Pakdil}}
\affiliation{Department of Physics and Astronomy, Dartmouth College, Hanover, NH 03755, USA}

\author{\OrcidIDName{0000-0001-6145-5859}{J.~Myles}}
\affiliation{Department of Astrophysical Sciences, Princeton University, Peyton Hall, Princeton, NJ 08544, USA}

\author{\OrcidIDName{0000-0001-9438-5228}{M. ~ Navabi}}
\affiliation{Department of Physics, University of Surrey, Guildford GU2 7XH, UK}

\author{\OrcidIDName{0000-0002-8282-469X}{N.~E.~D.~Noël}}
\affiliation{Department of Physics, University of Surrey, Guildford GU2 7XH, UK}

\author{\OrcidIDName{0000-0002-6021-8760}{A.~B.~Pace}}
\affiliation{Department of Astronomy, University of Virginia, 530 McCormick Road, Charlottesville, VA 22904, USA}

\author{\OrcidIDName{0000-0002-2762-2024}{A.~Porredon}}
\affiliation{Centro de Investigaciones Energ\'eticas, Medioambientales y Tecnol\'ogicas (CIEMAT), Madrid, Spain}

\author{\OrcidIDName{0000-0002-5933-5150}{J.~Prat}}
\affiliation{Nordita, KTH Royal Institute of Technology and Stockholm University, SE-106 91 Stockholm.}

\author{\OrcidIDName{0000-0002-7354-3802}{M.~Raveri}}
\affiliation{Department of Physics and INFN, University of Genova, Genova, Italy}

\author{\OrcidIDName{0000-0001-5805-5766}{A.~H.~Riley}}
\affiliation{Institute for Computational Cosmology, Department of Physics, Durham University, South Road, Durham DH1 3LE, UK}

\author{\OrcidIDName{0000-0001-9376-3135}{E.~S.~Rykoff}}
\affiliation{SLAC National Accelerator Laboratory, Menlo Park, CA 94025, USA}
\affiliation{Kavli Institute for Particle Astrophysics \& Cosmology, P.O.\ Box 2450, Stanford University, Stanford, CA 94305, USA}

\author{\OrcidIDName{0000-0002-1594-1466}{J.~D.~Sakowska}}
\affiliation{Department of Physics, University of Surrey, Guildford GU2 7XH, UK}

\author{\OrcidIDName{0000-0001-7147-8843}{S.~Samuroff}}
\affiliation{Institut de F\'{i}sica d'Altes Energies, The Barcelona Institute of Science and Technology, Campus UAB, 08193 Bellaterra (Barcelona) Spain}

\author{\OrcidIDName{0000-0003-3054-7907}{D.~Sanchez-Cid}}
\affiliation{Centro de Investigaciones Energéticas, Medioambientales y Tecnológicas (CIEMAT), Madrid, Spain}
\affiliation{Physik-Institut, University of Zurich, Winterthurerstrasse 190, CH-8057 Zurich, Switzerland}

\author{\OrcidIDName{0000-0003-4102-380X}{D.~J.~Sand}}
\affiliation{Steward Observatory, University of Arizona, 933 North Cherry Avenue, Tucson, AZ 85721-0065, USA}

\author{\OrcidIDName{0000-0003-3402-6164}{L.~Santana-Silva}}
\affiliation{Centro Brasileiro de Pesquisas F\'isicas, Rua Dr. Xavier Sigaud 150, 22290-180 Rio de Janeiro, RJ, Brazil}

\author{\OrcidIDName{0000-0002-1831-1953}{I.~Sevilla-Noarbe}}
\affiliation{Centro de Investigaciones Energ\'eticas, Medioambientales y Tecnol\'ogicas (CIEMAT), Madrid, Spain}

\author{\OrcidIDName{0000-0002-6389-5409}{T.~Shin}}
\affiliation{Department of Physics, Carnegie Mellon University, Pittsburgh, PA 15213}

\author{\OrcidIDName{0000-0001-6082-8529}{M.~Soares-Santos}}
\affiliation{Physik-Institut, University of Zurich, Winterthurerstrasse 190, CH-8057 Zurich, Switzerland}

\author{\OrcidIDName{0000-0003-1479-3059}{G.~S.~Stringfellow}}
\affiliation{Center for Astrophysics and Space Astronomy, University of Colorado, 389 UCB, Boulder, CO 80309-0389, USA}

\author{\OrcidIDName{0000-0001-7836-2261}{C.~To}}
\affiliation{Kavli Institute for Cosmological Physics, University of Chicago, Chicago, IL 60637, USA}

\author{\OrcidIDName{0009-0002-4207-0210}{A.~Tong}}
\affiliation{Department of Physics and Astronomy, University of Pennsylvania, Philadelphia, PA 19104, USA}

\author{\OrcidIDName{0000-0002-5622-5212}{M.~A.~Troxel}}
\affiliation{Department of Physics, Duke University Durham, NC 27708, USA}

\author{\OrcidIDName{0000-0003-4341-6172}{A.~K.~Vivas}}
\affiliation{Cerro Tololo Inter-American Observatory/NSF NOIRLab, Casilla 603, La Serena, Chile}

\author{\OrcidIDName{0000-0003-1585-997X}{M.~Yamamoto}}
\affiliation{Department of Astrophysical Sciences, Princeton University, Peyton Hall, Princeton, NJ 08544, USA}

\author{\OrcidIDName{0000-0002-9541-2678}{B.~Yanny}}
\affiliation{Fermi National Accelerator Laboratory, PO Box 500, Batavia, IL, 60510, USA}

\author{\OrcidIDName{0009-0006-5604-9980}{B.~Yin}}
\affiliation{Department of Physics, Duke University Durham, NC 27708, USA}

\author{\OrcidIDName{0000-0001-5969-4631}{Y.~Zhang}}
\affiliation{NSF National Optical-Infrared Astronomy Research Laboratory, 950 N Cherry Avenue, Tucson, AZ 85719}

\author{\OrcidIDName{0000-0001-9789-9646}{J.~Zuntz}}
\affiliation{Institute for Astronomy, University of Edinburgh, Edinburgh EH9 3HJ, UK}

\email{$^{\star}$dhayaa@uchicago.edu, chihway@kicp.uchicago.edu}

% Abstract of the paper
\begin{abstract}
We present cosmological constraints from the Dark Energy Camera All Data Everywhere (DECADE) cosmic shear analysis. This work uses shape measurements for $107$ million galaxies detected in Dark Energy Camera (DECam) imaging of $5,\!412$ deg$^2$ of the sky outside the Dark Energy Survey (DES) footprint. We derive constraints on the cosmological parameters $S_8 = 0.791^{+0.027}_{-0.032}$ and $\Om =0.269^{+0.034}_{-0.050}$ for the \LCDM model, which are consistent with those from other weak lensing surveys and from the cosmic microwave background. We combine our results with cosmic shear results from DES Y3 at the likelihood level, since the two datasets span independent areas on the sky. The combined measurements, which cover $\approx\! 10,\!000 \deg^2$, prefer $S_8 = 0.791 \pm 0.023$ and $\Om = 0.277^{+0.034}_{-0.046}$ under the \LCDM model. These results are the culmination of a series of rigorous studies that characterize and validate the \decade dataset and the associated analysis methodologies. Overall, the \decade project demonstrates that the cosmic shear analysis methods employed in Stage-III weak lensing surveys can provide robust cosmological constraints for fairly inhomogeneous datasets, where properties like seeing, survey depth, extinction, \etc vary significantly across the survey footprint. This opens the possibility of using data that have been previously categorized as ``unusable'' for cosmic shear analyses, thereby increasing the available area from upcoming weak lensing surveys by up to $\mathcal{O}(30\%)$.
\end{abstract}

%%%%%%%%%%%%%%%%% BODY OF PAPER %%%%%%%%%%%%%%%%%%

\section{Introduction}

Observational cosmology is now a mature field with a multitude of distinct observational probes, each of which provides unique insight into our understanding of the content and evolution of our Universe \citep[\eg][]{Allen:2011:Clusters, Goobar:2011:SNe, Mandelbaum:2018:WL, Planck:2020:Cosmo}. Weak (gravitational) lensing is one such cosmological probe that will be critical to further constrain the physics of the low-redshift Universe \citep{Spergel:2015:Roman, Euclid, LSST2018SRD} and also has the potential to improve our constraints on a wide variety of extended cosmological models such as modified gravity \citep[\eg][]{Schmidt:2008:MG_WL}, primordial signatures \citep[\eg][]{Anbajagane2023Inflation, Goldstein:2024:inflation, Primordial1, Primordial2}, \etc

Weak lensing is a phenomenon whereby light from distant sources is deflected by the presence of gravitational potentials --- sourced by the matter distribution --- between the sources and the observer \citep{Bartelmann2001, Schneider:2005:Lensing}. As a result, this phenomenon is sensitive to the distribution of all matter associated with this potential. Thus, weak lensing has long been regarded as an intrinsically clean cosmological probe that depends on only a few astrophysical processes, but is still sensitive to physics underlying the geometry of our Universe and to the growth of structure within it. A quarter-of-a-century after the first detection of this phenomenon \citep{Bacon2000, Wittman2000, Kaiser2000}, weak lensing has reached a state where the accuracy and robustness of the measurements (where the requirement of ``robustness'' is broadly categorized as systematics control) are as important as the statistical precision of the probe \citep{Asgari2021, Amon2021, Secco2021, Li2023}. This is especially true given apparent discrepancies between cosmological constraints from weak lensing in galaxy surveys and the constraints inferred from the cosmic microwave background (CMB) \citep[\eg][]{Asgari2021, Amon2021, Secco2021, Li2023, Planck:2020:Cosmo}. In particular, when assuming the $\Lambda$CDM model, weak lensing measurements prefer less structure in the matter distribution relative to the preference inferred from the CMB. This is usually referred to as the ``$\sigma_8$ tension'' or ``$S_8$ tension'', where $\sigma_8$ is the normalization of the linear, present-time matter power spectrum smoothed on $8\,h^{-1}$Mpc scales, and $S_{8}\equiv \sigma_{8}\sqrt{\Omega_{\rm m}/0.3}$, where $\Omega_{\rm m}$ is the ratio of the present-time matter energy density to the critical energy density.

In the context of these potential tensions, one of the most convincing cross-checks comes from performing independent analyses with different datasets, different algorithms, and preferably also different analysis teams. This is similar to the experimental design in high energy particle physics, where multiple groups --- \eg CMS \citep{CMS:2008xjf} and ATLAS \citep{ATLAS:2008xda}, or CDF \citep{Abe:1988me} and D\O\, \citep{Abachi:1994td} --- analyze particle collider data in a fully independent, blinded fashion. Thus, if a discovery is confirmed by multiple groups, it is unlikely to be a systematic introduced by a specific experimental design. The large-scale structure community did not intentionally develop such a format, but one naturally emerged with the three Stage-III\footnote{The ``Stage-X'' terminology was introduced in \citet{Albrecht2006} to describe the different phases of dark energy experiments. There are currently four stages, where Stage-III refers to the dark energy experiments that started in the 2010s and Stage-IV refers to those that start in the 2020s.} photometric galaxy surveys: the Dark Energy Survey \citep[DES,][]{Flaugher2005}, the Kilo-Degree Survey \citep[KiDS,][]{deJong2015} and the Hyper Suprime-Cam Subaru Strategic Program \citep[HSC-SSP,][]{Aihara2018}. To date, the headline weak lensing constraints from the three surveys give $S_8=0.759^{+0.025}_{-0.023}$ \citep[DES,][]{Secco2021, Amon2021}, $S_8=0.759^{+0.024}_{-0.021}$ \citep[KiDS,][]{Asgari2021} and $S_8=0.769^{+0.031}_{-0.034}$ \citep[HSC,][]{Li2023}. The combined, updated analysis of \citet{DESKiDS2023} gives $S_8=0.790^{+0.018}_{-0.014}$.

Interestingly, all these results find $S_8$ to be lower than the value inferred from the CMB \citep[$S_8=0.832 \pm 0.013$,][]{Planck:2020:Cosmo}. 
However, the regions of sky observed by three surveys --- DECADE, KiDS, and HSC --- partially overlap (see Figure \ref{fig:footprint}), so the information is not entirely independent. The robustness of the $S_8$ tension could increase if statistically independent measurements corroborate the low $S_8$ value, and vice versa --- if independent measurements were to find no evidence of tension. This work presents the cosmological constraints from a fourth weak lensing dataset, the Dark Energy Camera All Data Everywhere (\decade), that has similar constraining power as the aforementioned Stage-III surveys and is statistically independent from the DES dataset.

This paper is part of a series of works in the \decade cosmic shear project: \href{\#cite.paper1}{Anbajagane \& Chang et al. \citeyear{paper1}} (hereafter \citetalias{paper1}) describes the shape measurement pipeline, the derivation of the final source-galaxy sample for the weak lensing analysis, and the robustness tests and image simulation pipeline that characterize our measurements. \citet[][hereafter \citetalias{paper2}]{paper2} derives the tomographic bins and calibrated redshift distributions per bin for our source galaxy sample, together with a series of validation tests. \href{\#cite.paper1}{Anbajagane \& Chang et al. \citeyear{paper3}} (hereafter \citetalias{paper3}) describes the methodology and validation of our cosmological inference pipeline, in addition to a series of tests to evaluate the impact of survey inhomogeneity. Lastly, this work (\textsc{Paper IV}) shows our cosmic shear measurements and presents our constraints on parameters of different cosmological models.

The DECADE dataset is derived from multi-band Dark Energy Camera \citep[DECam,][]{Flaugher2015} imaging performed outside of the DES footprint. The catalog contains shape measurements for $107$ million galaxies assembled from $5{,}412$\,deg$^2$ of DECam imaging in the northern Galactic cap. The entire footprint is completely independent of DES. This dataset presents a unique opportunity to stress-test the $S_8$ tension in a number of ways. First, the \decade multi-band coadded images\footnote{These images are obtained by performing a weighted sum over all available camera exposures (``coaddition'') in a given part of the sky. See \citet{Morganson:2018}, \citet{Sevilla-Noarbe2021}, \citet{paper1}, \etc  for more details.} were derived using the image processing pipeline used in the DES Year 6 campaign \citep[DES Y6,][]{Bechtol:2025:Y6GOLD}. Second, the \decade source galaxy catalog uses the \textsc{Metacalibration} measurement algorithm \citep{Sheldon2017, Huff2017} to estimate lensing shear from galaxies, and the self-organizing map photometric redshift (SOMPZ) method \citep{Buchs2019, Myles:2021:DESY3, Sanchez:2023:highzY3} to calibrate the galaxy redshift distributions; this follows the choices in DES Y3 \citep{DESY3KP2022}. However, in the process of executing the \decade project, most parts of the analysis pipelines have been rewritten and tested, verifying their reliability.

Finally, the \decade catalog effectively doubles the sky coverage of precision weak lensing datasets. This improves the overlap between such weak lensing datasets and other wide-field cosmological surveys, including CMB experiments like \textit{Planck} \citep{Planck:2020:LegacyOverview}, the South Pole Telescope \citep[SPT,][]{Carlstrom2011}, the Atacama Cosmology Telescope \citep[ACT,][]{ACT:2007, ACT:2016}, and the Simons Observatory \citep[SO,][]{Simons:2019:Experiment}; spectroscopic datasets such as the Sloan Digital Sky Survey \citep[SDSS][]{York_2000, Dawson_2013, Dawson_2016} and the Dark Energy Spectroscopic Instrument \citep[DESI,][]{DESI:2016:Part1}; as well as X-ray surveys like eROSITA \citep{Merloni:2012:Erosita}. Future work will perform cross-correlation analyses using the \decade data to further stress-test the $S_8$ tension and to probe a variety of astrophysical and cosmological questions, as has already been pursued in the existing survey landscape \citep[\eg][]{Shin:2019:Splashback, Gatti2021DESxACT, Pandey2021DESxACT, Tilman:2022:tSZxWL, Chang2023, Omori:2023:CMBL, Sanchez:2023:tSZ, Anbajagane:2024:Shocks, Bigwood:2024:BaryonsWLkSZ}.

There is one more unique feature of the \decade catalog that will be highlighted throughout this paper: the \decade dataset was compiled from a collection of community-led DECam imaging campaigns and is significantly more inhomogeneous than the other Stage-III datasets mentioned above (\eg see Figure 12 in \citetalias{paper1} or Figure 1 in \citetalias{paper3}). Thus, the \decade analysis stress-tests many of the shear algorithms and models in a regime where the data quality is more variable. While the next-generation of dedicated lensing surveys, such as the Vera C.\ Rubin Observatory Legacy Survey of Space and Time \citep[LSST,][]{LSST:2009:ScienceBook}, the\textit{ Nancy Grace Roman Space Telescope} \citep{Spergel:2015:Roman}, and the \textit{Euclid} mission \citep{Euclid}, are expected to be more homogeneous than the \decade data, our findings still provide guidance on the level of variable data quality that can be accommodated in a lensing analysis. For example, the results of this work may motivate loosening data quality requirements when building a cosmology-ready sample, and thereby provide additional area/objects in the final sample of a given survey. This is especially relevant when considering the wide-fast-deep footprint of LSST, which could be extended to higher/lower Galactic latitudes \citep[\eg][]{Olsen:2018:BigSkyCadence}.

This work is structured as follows: in Section~\ref{sec:data_model}, we briefly describe the \decade dataset, the cosmological model, and parameter inference pipeline we use. In Section~\ref{sec:unblinding}, we summarize our blinding procedure, criteria for unblinding, and the results from the unblinding tests (a more detailed description is provided in Appendix \ref{appx:unblind_test}). The main results of this analysis are presented in Section~\ref{sec:Constraints}, and we discuss our findings in Section \ref{sec:Discussion}. We conclude in Section \ref{sec:summary}. Additional analysis results are shown in Appendix \ref{appx:DES_matched} and \ref{appx:more_param}, and we will refer to them in the main text where relevant.

\section{Data, modeling and inference}\label{sec:data_model}

We briefly summarize the \decade dataset, our modeling choices, and our approach to parameter inference. For more technical details and discussion, we will direct the reader to results and discussions from the other papers in this series (\citetalias{paper1}, \citetalias{paper2}, \citetalias{paper3}).

\subsection{Data}\label{sec:data_model:data}

\begin{figure*}
    \includegraphics[width = 2\columnwidth]{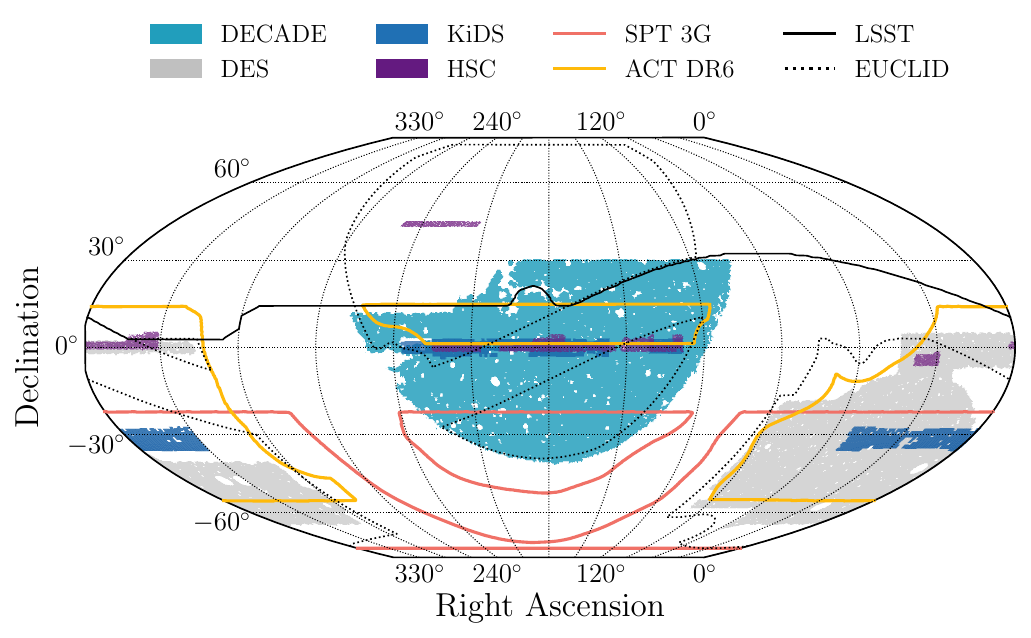}
    \caption{The footprint of the \decade cosmic shear analysis, in relation to those of three other Stage-III surveys: DES Y3 (grey), KiDS-1000 (dark blue), and HSC Y3 (purple). We also show the footprints for the LSST wide-field survey (black solid), the \textit{Euclid} wide-field survey (black dotted), the SPT Ext-10k survey (orange), and ACT DR6 (yellow). See the introduction for references to the different experiments.} 
    \label{fig:footprint}
\end{figure*}

\begin{figure}
    \centering
    \includegraphics[width = \columnwidth]{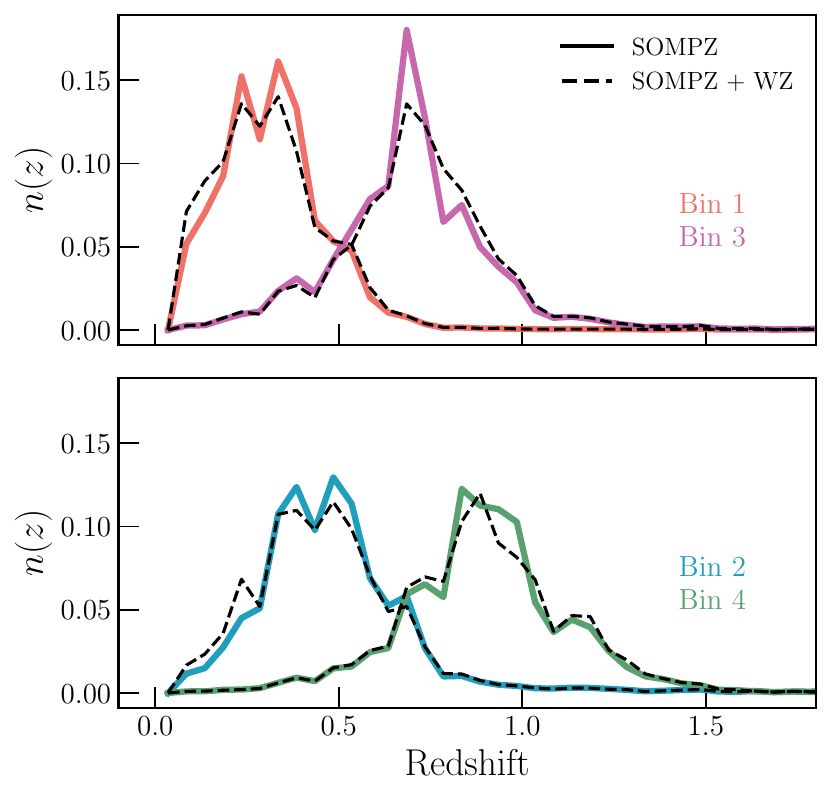}
    \caption{Redshift distributions, $n(z)$, for the four tomographic bins defined in the source galaxy sample. We show both the fiducial redshift distributions derived from the SOMPZ method and also those from combining SOMPZ with information from the clustering redshift (WZ) method \citepalias[see Section 3.7 of][]{paper2}.}
    \label{fig:nofz}
\end{figure}

The \decade dataset is a galaxy shape catalog of 107 million galaxies spanning $5,\!412 \deg^2$ of the sky; see Figure \ref{fig:footprint} for the survey footprint compared to those of other surveys. The catalog is introduced in \citetalias{paper1}, alongside (i) a suite of null-tests that validate the absence of statistically significant systematics, and; (ii) an independent test/calibration of the measurement pipelines using an end-to-end image simulation campaign. The shear $\gamma_{1,2}$ is a distortion to the shape and orientation of an object due to weak lensing. It is estimated for each galaxy using the \textsc{Metacalibration} method \citep{Sheldon2017, Huff2017}, with an approach designed to mimic that of DES Y3 \citep{y3-shapecatalog, Macrann2022ImSim}. In \citetalias{paper2}, we split the shape catalog into four tomographic bins, and then estimate the redshift distribution of the ensemble in each bin using self-organizing maps \citep[SOMPZ,][]{Buchs2019, Myles:2021:DESY3}. We also cross-check these redshift estimates using a spatial clustering-based approach \citep[WZ, ][]{Menard2013, Davis2017, Gatti:2022:WzY3}. Figure~\ref{fig:nofz} shows the fiducial redshift distribution of our sample, as well as the distribution obtained from combining SOMPZ and the clustering-based approach.

\begin{table}
    \centering
    \begin{tabular}{c|c|c|c|c|c}
     \hline
      & $n_{\rm gal}$ & $R_{\rm tot, 1}$ & $R_{\rm tot, 2}$ & $n_{\rm eff, H12}$ & $\sigma_{e, {\rm H12}}$ \\
     \hline
    Bin 1 & 1.396 & 0.836 & 0.837 & 1.239 & 0.233 \\
    Bin 2 & 1.371 & 0.771 & 0.771 & 1.150 & 0.259 \\
    Bin 3 & 1.375 & 0.740 & 0.742 & 1.169 & 0.248 \\
    Bin 4 & 1.369 & 0.620 & 0.621 & 1.153 & 0.289 \\
    Full sample & 5.511 & 0.756 & 0.757 & 4.586 & 0.254 \\
        \hline
    \end{tabular}
    \caption{The raw source galaxy number density ($n_{\rm gal}$), different components of the total shear response ($R_{{\rm tot}, 1/2}$), effective number density of weak lensing galaxies ($n_{\rm eff}$) and shape noise ($\sigma_{e}$) in the \citet{Heymans2012} definition (denoted as H12), for each of the tomographic bins as well as the full non-tomographic sample. The source galaxy number densities are calculated with an area of 5,412 deg$^2$, and are presented in units of arcmin$^{-2}$.}
    \label{tab:neff_w2}
\end{table}

The \decade dataset is derived by combining available DECam community imaging data in the northern Galactic cap; see Figure \ref{fig:footprint} for the survey footprint. The nature of this dataset --- as an amalgamation of available archival data, rather than as a dedicated weak lensing survey program --- results in significant inhomogeneities in the survey observing conditions such as exposure time and image quality. This can propagate into variations in the observed (noisy) object properties, and therefore the detection/selection functions of galaxy samples. While inhomogeneities in observing conditions exist for other galaxy weak lensing surveys, such as DES, the amplitude of these variations are smaller than those found in the \decade survey; see Figure 1 in \citetalias{paper3} for a comparison. We have performed a variety of tests quantifying the impact of such inhomogeneities on cosmology constraints, and have found no evidence that these effects are larger than the precision level of our data \citepalias[see Section 6 of][]{paper3}. A summary of the dataset, including the source galaxy number density, $n$, and shape noise per tomographic bin, $\sigma_e$, are found in Table \ref{tab:neff_w2}  (reproduced from Table 2 of \citetalias{paper1}).

Our fiducial cosmology constraints are derived from measurements of the shear two-point correlation functions. Following \citet{Bartelmann2001} and \citet{Schneider:2005:Lensing}
\begin{equation}\label{eqn:2pt_def}
    \xi_\pm = \langle \gamma_t^a \gamma_t^b \rangle_{ab} \pm \langle \gamma_\times^a \gamma_\times^b \rangle_{ab},
\end{equation}
where the averages are taken over all galaxy pairs $a,b$, and $\gamma^{a,b}_t$ and $\gamma_\times^{ab}$ are the tangential and cross components of the shear, respectively, as defined with respect to the projected angular separation between galaxies $a$ and $b$ \citep[\eg see Equation 17 in][]{Schneider:2005:Lensing}.

\begin{figure*}
    \centering
    \includegraphics[width = 1.99\columnwidth]{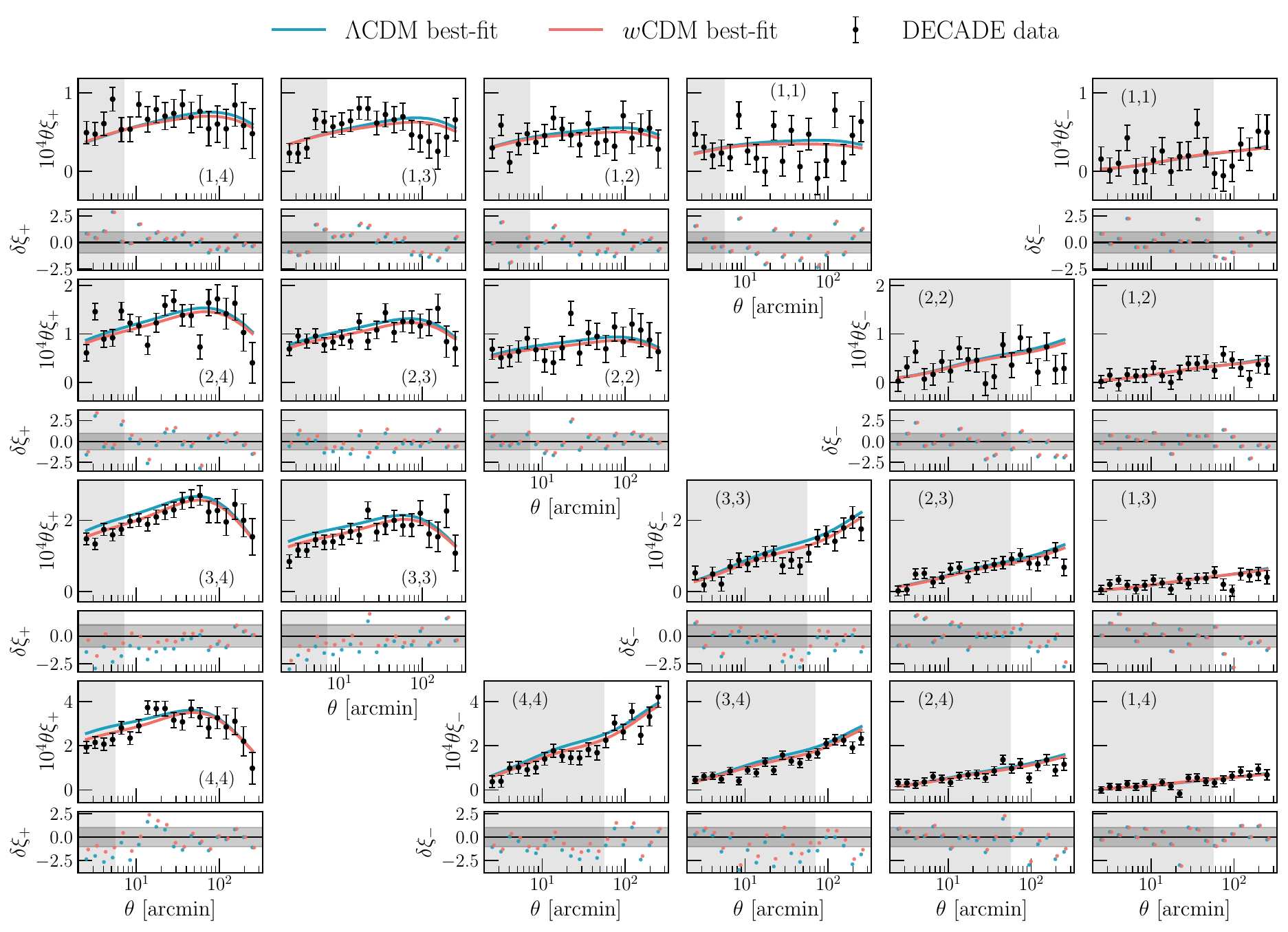}
    \caption{Our cosmic shear data vector and best-fit predictions under a \LCDM and wCDM model. The upper (lower) triangle shows the $\xi_+$ ($\xi_-$) data vectors. Each panel corresponds to a bin pair, denoted in the legend. The shaded region marks angular scales removed during inference due to uncertainty in modeling baryonic effects in the matter power spectra. We show normalized residuals, $\delta \xi_\pm = (\xi^{\rm\, data}_\pm - \xi^{\rm\, theory}_\pm)/\sigma_{\xi_\pm}$, in the lower sub-panel for each bin pair. The horizontal gray band covers the 1$\sigma$ region around $\delta \xi_\pm = 0$.}
    \label{fig:datavec}
\end{figure*}

We do not have access to the shear field at the location of each galaxy, and therefore obtain it by averaging over the \textit{ellipticities}, $e$, of many galaxies. In practice, we estimate $\xi_\pm$ as,
\begin{equation}\label{eqn:2pt_meas}
    \xi_{\pm}^{ij}(\theta) = \frac{\sum_{ab} w_a w_b \left( \hat{e}^i_{t,a} \hat{e}^j_{t,b} \pm \hat{e}^i_{\times,a} \hat{e}^j_{\times,b} \right)}
     {\langle R\rangle_a \langle R\rangle_b\sum_{ab} w_a w_b}, |\vec{\theta_a} - \vec{\theta_b}| \in \theta,
\end{equation}
where $i,j$ are indices over the four tomographic bins, $\hat{e}$ are the mean-subtracted ellipticity estimates from \textsc{Metacalibration}, $w$ are the inverse-variance weights defined in \citetalias{paper1}, and $\langle R \rangle$ is the weighted-average of the response function measured from \textsc{Metacalibration} (see Table \ref{tab:neff_w2} and \citetalias{paper1} for more details). The term $|\vec{\theta_a} - \vec{\theta_b}| \in \theta$ specifies that the sum $\sum_{ab}$ is only over galaxy pairs with angular separations in the angular bin $\theta$. The measured data vector is shown in Figure \ref{fig:datavec}, alongside the best-fit models derived from the analysis in Section \ref{sec:Constraints}. We discuss the data vector further in that section.

There have also been extensive studies in the lensing community of alternative summary statistics beyond the two-point correlation functions, and the corresponding constraining power on the parameters of interest \citep[\eg][]{Jain:1998:HOS, Friedrich:2018:DensitySplit, Gatti:2022:Moments, Secco2022MassAp, Euclid:2023:HOS, Anbajagane:2023:CDFs, Gatti:2024:WPHSims}. While we follow the standard convention of the lensing community in this work and use the extensively tested two-point correlation functions as our measurement, we note that many works have performed robust lensing analyses using these alternative statistics \citep[\eg][]{Gruen:2018:DensitySplit, Zucher:2022:Peaks, Fluri:2022:KiDS, Gatti:2022:Moments, Gatti:2024:WPHSims, Jeffrey:2025:Likelihood, Cheng:2025:WPH}. Such studies on the \decade data will be pursued in the future.

\subsection{Modeling and inference}\label{sec:data_model:model}

A detailed description of our modeling choices are found in \citetalias{paper3}. In brief, our modeling pipeline follows that of DES Y3 \citep{Krause2021}. The only difference is our choice to use \textsc{HMCode} \citep{Mead2020a} as our model for the nonlinear matter power spectrum, whereas DES Y3 used \textsc{HaloFit} \citep{Takahashi2012}. The former has been shown to be more accurately match measurements from simulations when compared to the latter \citep{Mead2016, Mead2020a, Mead2021b} and was used in the DES \& KiDS joint analysis \citep{DESKiDS2023}.

Given weak lensing (or shear) is sourced by the matter distribution, the shear two-point correlations can be predicted from the matter two-point correlations, \ie from the matter power spectrum. In particular, the shear two-point correlation functions can be modeled as,
\begin{align} \label{eq:xipm}
    \xi_{\pm}^{ij}(\theta)= \sum_{\ell} & \,\,\frac{2\ell+1}{2\pi\ell^{2}\left(\ell+1\right)^{2}}\left[G_{\ell,2}^{+}\left(\cos\theta\right)\pm G_{\ell,2}^{-}\left(\cos\theta\right)\right]\nonumber\\& \times\left[C_{EE}^{ij}(\ell)\pm C_{BB}^{ij}(\ell)\right],
\end{align}
where $\ell$ is the harmonic multipole and the functions $G^\pm_\ell(x)$ are computed from Legendre polynomials $P_\ell(x)$ and averaged over angular bins \citep{Krause2021}. The $i$ and $j$ indices specify the two tomographic redshift bins from which the correlation function is calculated. The term $C_{EE}$ is the angular matter power spectrum integrated along the line-of-sight after being weighted by the lensing kernels (see Equation 2 in \citetalias{paper3}).

The intrinsic alignments (IA) of galaxies also contributes to the $C_{EE}$ and $C_{BB}$ terms \citep[\eg][]{TroxelIshak2014, Lamman:2024:IA}. The IA signal is also connected to the matter distribution and therefore can be predicted using the matter power spectrum. These IA signal are included in our predictions using the Tidally Aligned Tidally Torqued \citep[TATT,][]{Blazek2019} model, following \citet{Secco2021} and \citet{Amon2021}. Similar to the latter two works, we also test a few variants for the IA model choice, including one with the simpler, nonlinear linear alignment model \citep[NLA,][]{Bridle2007} and one with no IA. The amplitude of the IA contribution is parameterized as,
\begin{align}\label{eqn:IA_ampl}
    A_1(z) &= -a_1 \bar{C}_1 \frac{\rho_{\text{crit}} \Om}{D(z)} 
    \left( \frac{1 + z}{1 + z_0} \right)^{\eta_1}, \\
    A_2(z) &= 5a_2 \bar{C}_1 \frac{\rho_{\text{crit}} \Om}{D^2(z)} 
    \left( \frac{1 + z}{1 + z_0} \right)^{\eta_2},\\
    A_{1\delta}(z) &= b_{\rm TA} A_1(z),
\end{align}
where $A_1$ and $A_2$ scale the matter power spectra, $D(z)$ is the linear growth rate, $\rho_{\text{crit}}$ is the critical density at $z = 0$, and $\bar{C}_1 = 5 \times 10^{-14} M_\odot h^{-2} \text{Mpc}^2$ is a normalization constant, set by convention. We choose a pivot redshift, $z_0 = 0.62$ following \citet{Secco2021, Amon2021}. The free parameters of our model are the amplitudes $a_1, a_2, b_{\rm TA}$ and the power-law indices $\eta_1, \eta_2$. The NLA model is obtained by setting $a_2, \eta_2, b_{\rm TA} = 0$.
For clarity, we reproduce Equations 20--23 from \citet{Secco2021}, which list the different IA-related power spectra that contribute to the final signal,
\begin{align}
\label{eq:dEtot}
P_{\mathrm{GI}}(k) =&\,\, A_1 P_{\delta}(k) + A_{1\delta} P_{0|0E}(k)
+ A_2 P_{0|E2}(k)~, \\[5pt]
\label{eq:EEtot}
P_{\mathrm{II},\mathrm{EE}}(k) =&\,\,
A_1^2 P_{\delta}(k) +
2 A_1 A_{1\delta} P_{0|0E}(k)
+ A_{1\delta}^2 P_{0E|0E}(k)\notag\\
&+ A_2^2 P_{E2|E2}(k)
+ 2 A_1 A_2 P_{0|E2}(k)\\
&+ 2 A_{1\delta} A_2 P_{0E|E2}(k)~,\notag\\[5pt]
\label{eq:BBtot}
P_{\mathrm{II},\mathrm{BB}}(k) =&\,\, A_{1\delta}^2 P_{0B|0B}(k)
+ A_2^2 P_{B2|B2}(k)\notag\\
& + 2 A_{1\delta} A_2 P_{0B|B2}(k) ~.
\end{align} 
The full expressions, and derivations, for each power spectra are detailed in \citet{Blazek2019}. See their Equations 37-39, and Appendix A in particular. We do not reproduce those expressions here and direct readers to \citet{Blazek2019} for technical details on the IA calculations. The terms $P_{\mathrm{GI}}(k)$ and $P_{\mathrm{II},\mathrm{EE}}(k)$ contribute to the angular power spectra $C_{EE}(\ell)$---in addition to the shear-shear correlation, which is of primary interest in this work---while $P_{\mathrm{II},\mathrm{BB}}(k)$ is the sole contribution to $C_{BB}(\ell)$.

\begin{table}
    \centering
    \begin{tabular}{l l}
        Parameter & Prior \\
        \hline
        $\Omega_{\rm m}$ &  $\mathcal{U}(0.1,0.9)$ \\
        $\Omega_{b}$ &  $\mathcal{U}(0.03,0.07)$\\
        $h$ &  $\mathcal{U}(0.55,0.91)$ \\
        $A_s \times10^9$ &  $\mathcal{U}(0.5,5)$ \\
        $n_s$ & $\mathcal{U}(0.87,1.07)$ \\
        $\Omega_{\nu} h^2$ &  $\mathcal{U}(0.0006, 0.00644)$  \\
        \hline
        $a_{1}$ &  $\mathcal{U}(-4,4)$ \\
        $a_{2}$ &  $\mathcal{U}(-4,4)$ \\
        $\eta_{1}$ & $\mathcal{U}(-4,4)$\\
        $\eta_{2}$ & $\mathcal{U}(-4,4)$\\
        $b_{\rm ta}$  & $\mathcal{U}(0,2)$ \\       
        \hline
        $\Delta z_1$  &$\mathcal{N}(0, 0.0163)$ \\
        $\Delta z_2$  &$\mathcal{N}(0, 0.0139)$ \\
        $\Delta z_3$  &$\mathcal{N}(0, 0.0101)$ \\
        $\Delta z_4$  &$\mathcal{N}(0, 0.0117)$ \\
        \hline
        $m_1$ & $\mathcal{N}(-0.00923, 0.00296)$  \\
        $m_2$ & $\mathcal{N}(-0.01895, 0.00421)$ \\
        $m_3$ & $\mathcal{N}(-0.04004, 0.00428)$ \\
        $m_4$ & $\mathcal{N}(-0.03733, 0.00462)$ \\
        \hline
    \end{tabular}
    \caption{Cosmological and nuisance parameters in the baseline \LCDM model. Uniform distributions in the range $[a,b]$ are denoted $\mathcal{U}(a,b)$ and Gaussian distributions with mean $\mu$ and standard deviation $\sigma$ are denoted $\mathcal{N}(\mu,\sigma)$. }
    \label{tab:params}
\end{table}

As discussed in \citetalias{paper3}, we fit the model above to our $\xi_{\pm}$ measurements using a Markov Chain Monte Carlo (MCMC) approach. We assume a Gaussian likelihood $L$, with 
\begin{equation} \label{eq:likelihood} 
    \ln L ( \xi_{\pm,d} | \boldsymbol{p})
    = -\frac{1}{2}\bigg(\xi_{\pm,d} - \xi_{\pm,m}(\mathbf{p})\bigg)\mathbf{C}^{-1} \bigg(\xi_{\pm,d} - \xi_{\pm,m}(\mathbf{p})\bigg), 
\end{equation} 
where $\xi_{\pm}$ is a concatenation of the $\xi_+$ and $\xi_-$ measurements; $\xi_{\pm,d}$ and $\xi_{\pm,m}$ are the data vectors measured in the data (d) and predicted from our theoretical model (m);$\textbf{C}^{-1}$ is the inverse covariance of the measurements; $\textbf{p}$ is a vector of the cosmology parameters and nuisance parameters listed in Table~\ref{tab:params}. The Bayesian posterior is given via the product of the likelihood $L$ and the prior $P$, or 
\begin{equation} 
    P(\mathbf{p}|\xi_{\pm,d}, M) =  \frac{\mathcal{L} ( \xi_{\pm,d} | \mathbf{p}, M)P(\mathbf{p} | M)}{P( \xi_{\pm,d} | M)},
\end{equation} 
where the denominator, $P( \xi_{\pm,d} | M)$, is the evidence of the data. The entire expression is conditioned on a model choice, $M$.
Table~\ref{tab:params} lists the priors and fiducial values of all the model parameters. Our covariance matrix is generated using \textsc{CosmoCov} \citep{Krause:2017:CosmoLike, Fang2020, Fang2020b} and follows the model of \citet{Friedrich:2021:CovY3} as used in DES Y3. This includes a simple Gaussian covariance, as well as a connected four-point term to account for nonlinear structure \citep{Wagner:2015:Response, Barreira:2017:Response, Barreira:2017:ResponseCov}, a super-sample contribution to incorporate correlations between small-scale modes as generated by modes larger than the survey footprint \citep[\eg][]{Barreira:2018:SSC}, and also a correction for the impact of the survey mask on the shape noise term \citep{Troxel:2018:Cov}.

All parameter inference is performed using the \textsc{CosmoSIS} package \citep{Zuntz2015}. We
use \texttt{Nautilus} \citep{Lange:2023:Nautilus} as the MCMC sampler for our fiducial chains, but we have checked that our results are consistent if we use the \texttt{Polychord} sampler \citep{Handley:2015:Polychord}, which is the default choice in DES. The exact hyper-parameters used for the two samplers are listed in Table 2 of \citetalias{paper3}.

Finally, in addition to our fiducial constraints, we also present joint constraints from \decade and DES Y3. Given that the \decade data comes from an independent patch of the sky relative to DES, there is negligible correlation between the two measurements and we can perform a joint analysis by simply multiplying the likelihoods from the two surveys. When doing so, there are two possible approaches for the IA model choices: first, we can use an independent set of parameters for each survey, resulting in 10 (4) free parameters under the TATT (NLA) model. Second, we can use a common set of parameters for both surveys, resulting in 5 (2) free parameters under the TATT (NLA) model. The latter choice is motivated by the fact that the galaxy selection (and detection) function is similar between \decade and DES Y3 given the number of common choices across the image processing and catalog definition. We use the former, more conservative choice for our fiducial analysis and present the latter in Appendix \ref{appx:IA}. We also use the fiducial scale cuts from \decade and DES Y3 for this analysis, and have verified that the contamination from baryons causes a $<0.3\sigma$ shift in the $\Omega_{\rm m}$ - $S_8$ plane of the joint analysis. Thus, the existing scale cuts pass the criteria defined by both \decade \citepalias{paper3} and DES Y3 \citep{Secco2021, Amon2021}.

\section{Blinding and unblinding}
\label{sec:unblinding}

Modern cosmology analyses using galaxy surveys typically implement a \textit{blinding} procedure \citep{Heymans2021,DESY3KP2022,Sugiyama2023}, with the goal of preventing analysis choices from implicitly biasing the cosmological results. Under a blinded analysis, all necessary checks of (and potential modifications to) the analysis pipeline are done without knowledge of the final cosmology constraints. Through this setup, we ensure we do not iteratively modify our analysis pipeline until we obtain a cosmology result deemed favorable in some manner. Different experiments have different approaches in blinding and unblinding. Overall, all Stage-III experiments have had fairly successful blinding experiences with minimal-to-no analysis changes made post-unblinding. This fact also makes the overall agreement between the different experiments' constraints significant and highly nontrivial.

In this work, we apply blinding to our data vector according to the methodology described in \citet{Muir2020}. In brief, we have a fiducial cosmology --- that is a set of point values for the cosmological parameters --- and an alternative one (chosen at random) that we use for blinding. We compute simulated data vectors ($\xi_{\pm}$) for both cosmologies and find the shift induced by changing from one cosmology to another. This shift is then added to the measured data vector. In this work, our fiducial cosmology is given in Table 1 of \citetalias{paper3}, and our alternative cosmology is given by taking the same and replacing the two cosmology parameters $(\sigma_{8}, w)$ with random values drawn from a uniform distribution of $([0.714, 0.954],[-1.5,-0.5])$. Under this approach, our model will still be able to fit the blinded data vector with a reasonable $\chi^2$.

Note that our galaxy shape catalogs were not blinded, unlike what is done in DES Y3 and Y6 analyses \citep{y3-shapecatalog, Yamamoto2025}. Given the uniqueness of our dataset --- and the potential oddities that could consequently arise and require detailed studies --- we chose to keep the shape catalog unblinded so as to retain more flexibility in our analysis in the event of such scenarios. Furthermore, the catalog-level tests (those described in \citetalias{paper1}) do not use/show measurements that are directly connected to the cosmological inference. Given the relatively small size of the DECADE team, safeguards against inadvertent unblinding (\eg by comparing our $\xi_\pm$ data vector with DES) were considered sufficient. Appendix~\ref{appx:unblind_test} describes the full set of tests that we passed before unblinding.

\section{Results}\label{sec:Constraints}

This section presents all main results from the \decade cosmic shear analysis. First, Figure~\ref{fig:datavec} shows the measured data vectors in comparison with the best-fit $\Lambda$CDM and $w$CDM models. Both models are good descriptions of the data, with $\chi^2/N_{\rm data}= 264.7/220$ ($\Lambda$CDM) and $\chi^2/N_{\rm data}= 264.26/220$ ($w$CDM), where $N_{\rm data}$ is the number of data points. The effective number of constrained parameters\footnote{Calculated via $N_{\rm eff} \equiv N -{\rm Tr}(C_\Pi^{-1} C_p)$, where $N$ is the number of free parameters in the model, $C_p$ is the posterior covariance and $C_\Pi$ is the prior covariance.} are 6.73 and 7.15 for the two models, which include 2.02 and 2.48 cosmological parameters, respectively.  

We note that for the rest of this work we quote constraints from using the \texttt{Nautilus} sampler \citep{Lange:2023:Nautilus}, but we have checked that the numerical values we quote (mean, standard deviations, maximum-likelihood) are consistent with those estimated from \texttt{Polychord} \citep{Handley:2015:Polychord}, which is the standard sampler in DES Y3. Line 2 in Figure~\ref{fig:data_variation} below shows this comparison. We chose to use \texttt{Nautilus} because its importance sampling scheme requires significantly fewer likelihood evaluations to provide a converged chain. 

Finally, all constraints quoted below show the mean and the 68\% confidence interval, following DES Y3 \citep[][]{DESY3KP2022}. The confidence interval is computed using the 16\% and 84\% values of the distribution. All estimates of ``distance'' between two posteriors are calculated using a simple metric, using $\Seight$ alone; see Equation \ref{eq:s8_dist}. The posteriors in the cosmological parameters are always in good agreement with each other, and therefore we opt for this simpler estimator rather than the sophisticated metrics used in, for example, DES Y3 \citep{Doux:2021:tensions, Raveri:2021:NGTension}.

\subsection{$\bold{\Lambda}$CDM}
\label{sec:lcdm}

\begin{figure*}
    \centering
    \includegraphics[width = 0.9\columnwidth]{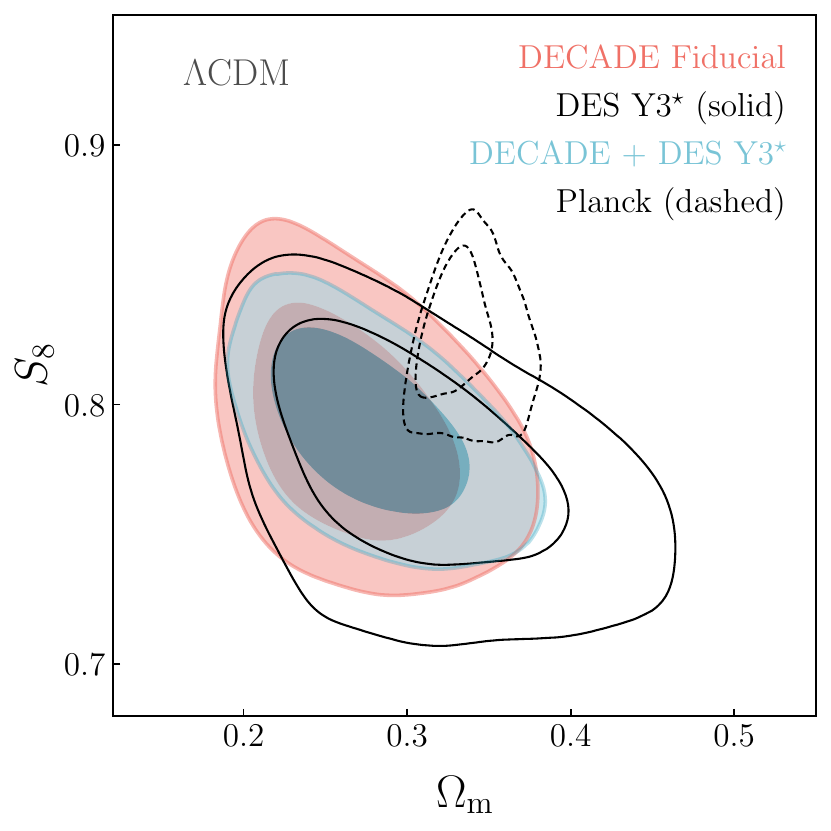}\hspace{30pt}
    \includegraphics[width = 0.9\columnwidth]{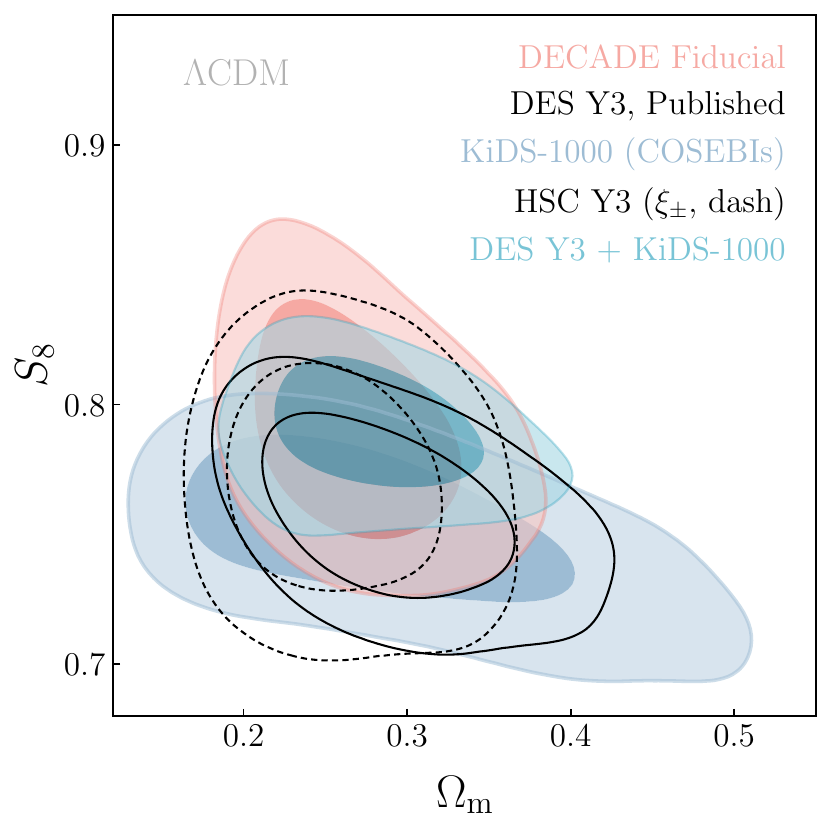}
    \caption{The fiducial \LCDM cosmic shear constraints for $\Omega_{\rm m}$ and $S_8$. The $\star$ denotes constraints reanalysed using our pipeline. All constraints are also listed in Table \ref{tab:constraints}. \textbf{Left:} Results from \decade (red), DES Y3 analyzed with our pipeline (black solid), the combination of the two (blue), and \textit{Planck} (black dashed). The \decade data constrains $S_8=0.791^{+0.027}_{-0.032}$, which is consistent with DES Y3 and also has similar constraining power to DES Y3 when both are analyzed with the same pipeline. The $S_8$ constraints from \decade (\decade + DES Y3) are consistent with \textit{Planck} within the $1.1\sigma$ ($1.2\sigma$) level. \textbf{Right:} Comparison of the \decade constraints with other published results. We do not reanalyze any data in this case and use the public posteriors. The DECADE results are consistent with existing constraints within $\lesssim 1 \sigma$. See Section \ref{sec:sec:Consistency} for references and links to the chains from the other lensing surveys.}
    \label{fig:constraints}
\end{figure*}

We start by constraining the \LCDM model. Figure~\ref{fig:constraints} shows the posterior constraints for $\Seight$ and $\Om$, while the posterior (and prior) of the full parameter space is shown in Appendix~\ref{appx:fullparam}. Under a \LCDM model, our constraints from \decade are
\begin{align}
    \Seight & =  0.791^{+0.027}_{-0.032} \\[5pt]
    \Om & =  0.269^{+0.034}_{-0.050}.
\end{align}
Figure~\ref{fig:constraints} also overlays the constraints from DES Y3 obtained after reanalyzing the data vector using our inference pipeline. This differs from the analysis of \citet{Secco2021} and \citet{Amon2021} in not using information from DES shear ratios \citep{Sanchez2022} and in switching to an updated model for the nonlinear matter power spectrum; see Appendix \ref{appx:DES_matched} for details and comparison tests. We denote reanalyzed constraints with $\star$ for clarity. The \decade and DES Y3 data, both analyzed with our pipeline, are consistent within the $0.3\sigma$ level. We use Equation~\ref{eq:s8_dist} to calculate these differences.

The same figure also shows constraints from \textit{Planck} using their ``TT+TE+EE+LowE'' measurements \citep{Planck:2020:Cosmo}. We extract the corresponding posteriors using the public \textit{Planck} likelihood\footnote{\url{https://wiki.cosmos.esa.int/planck-legacy-archive/index.php/CMB_spectrum_\%26_Likelihood_Code}} but using priors on cosmological parameters as shown in Table~\ref{tab:params}.\footnote{In addition to the parameters in Table \ref{tab:params} we also vary the optical depth $\tau$ with a uniform prior between 0.01 and 0.8.} The $S_8$ constraints from \decade and \textit{Planck} are consistent within the $1.1\sigma$ level. The constrained values of the cosmological parameters are listed in Table \ref{tab:constraints} below.

The results of Figure~\ref{fig:constraints} show the $S_8$ constraint from \decade has similar precision to that from the reanalyzed DES Y3 data, while the $\Om$ constraint is notably more precise in the former compared to the latter. However, our simulated analysis finds the two surveys' constraints have similar precision on $\Om$. The difference in the data constraints on $\Om$ may be due to mild degeneracies between $\Om$ and the IA amplitudes ($a_1, a_2$). This degeneracy is alleviated slightly in \decade, relative to DES Y3, as the IA amplitudes of the former (particularly $a_1$) are well-constrained (Figure \ref{fig:IA}).

\subsubsection{Combined constraints from DECADE and DES Y3}

The left panel of Figure~\ref{fig:constraints} shows the combined constraints of DES Y3 and \decade (blue), obtained using the \decade pipeline. We reiterate that the two surveys can be trivially combined because (1) the constraints are consistent with each other, and; (2) the datasets are entirely independent on the sky and so their measurements will have negligible cross-correlations. Therefore, the two surveys can be combined at the likelihood level. As a reminder (see Section \ref{sec:data_model:model} for more details), each dataset has its own set of shear calibration ($m$) and redshift calibration ($\Delta z$) parameters. While we perform two versions of the joint constraints --- where the IA parameters are independent/shared for each dataset --- we only show the constraint from the independent IA case and relegate the latter to Appendix \ref{appx:IA}. When using independent IA parameters, we find:
\begin{align}
    \Seight & =  0.791\pm 0.023 \\
    \Om & =  0.277^{+0.034}_{-0.046}
\end{align}
The joint constraint on $\Seight$ is consistent with that of \textit{Planck} at the $1.2\sigma$ level.

The precision on $S_8$ after combining \decade and DES Y3 is $26\%$ better than those of the individual surveys. Given the two weak lensing datasets have similar constraining power, the combination of the two can be expected to reduce statistical uncertainties (relative to the single-survey case) by a factor of $1/\sqrt{2} \approx 29\%$.\footnote{We do not expect any degeneracy breaking --- which would increase the expected improvement --- when combining two similar datasets from the same cosmological probe.} Our simulated tests confirm the uncertainties on $S_8$ in the combined analysis are 29\% smaller than those from the single-survey analyses. In practice, the \decade constraint on $S_8$ is slightly better than that from DES Y3 (see Table \ref{tab:constraints}). We therefore expect the improvement in the combined analysis to be slightly lower than our theoretical expectation, as is found in the case above.

Our combined constraints on $S_8$ are weaker than those from combining DES Y3 and KiDS-1000, quoted by \citet{DESKiDS2023} to be $\Seight = 0.790^{+0.018}_{-0.014}$. This is because the scale cuts used in our analysis are more conservative than those used in the latter work, and our choice of using the TATT IA model is also more flexible than the NLA model used in the latter. Section \ref{sec:sec:IA} discusses the improvement in the \decade-only constraints if we use the NLA model or use all scales in the analysis. Also, note that the Figure-of-Merit (FoM) --- defined using the 2D parameter posterior covariance as $(\det C(\Seight, \Om))^{-1/2}$ and listed in Table \ref{tab:constraints} --- of the \decade and DES Y3$^\star$ combined analysis is only 12\% lower than that of the DES Y3 and KiDS-1000 combined analysis.

\subsubsection{Consistency with other cosmic shear surveys}\label{sec:sec:Consistency}

In the right panel of Figure~\ref{fig:constraints}, we compare constraints from \decade with other Stage-III cosmic shear surveys, as published by each of the collaborations. We do not reanalyze their data vectors with our own pipeline. Note that in this case, each survey makes assumptions and assigns priors (for both cosmology and nuisance parameters) that differ from those of other surveys. While such choices do not change the qualitative findings of each survey, they still somewhat shift the exact constraints \citep[\eg][]{Chang:2019:Reanalysis, Longley2023, DESKiDS2023}. With that caveat, we find that all of the recent cosmic shear constraints from Stage-III surveys are consistent with each other. We show results from DES Y3\footnote{\url{https://desdr-server.ncsa.illinois.edu/despublic/y3a2_files/chains/chain_1x2pt_lcdm_SR_maglim.txt}} \citep{Secco2021, Amon2021}, KiDS-1000\footnote{\url{https://kids.strw.leidenuniv.nl/DR4/data_files/KiDS1000_cosmic_shear_data_release.tgz}} \citep{Asgari2021}, HSC Y3\footnote{\url{https://idark.ipmu.jp/~xiangchong.li/HSC/HSCY3/Li2023/hsc_y3_real_cosmic_shear.txt}} \citep{Li2023, Dalal2023}, and then the combination of the DES Y3 and KiDS-1000 data\footnote{\url{https://desdr-server.ncsa.illinois.edu/despublic/y3a2_files/y3a2_joint-des-kids/chains/chain_desy3_and_kids1000_hybrid_analysis.txt}} \citep{DESKiDS2023}. The KiDS-1000 analysis of \citet{Asgari2021} present constraints from analysing multiple different data vector, and we show constraints from the COSEBIs data vector, while HSC Y3 has a real-space \citep{Li2023} and harmonic-space \citep{Dalal2023} analysis, of which we show the former. The constraints from the companion analyses/data vectors that are omitted in Figure \ref{fig:constraints} are instead listed in Table \ref{tab:constraints}.

\subsection{$\bold{\rm w}$CDM}

\begin{figure*}
    \centering
    \includegraphics[width = 1.5\columnwidth]{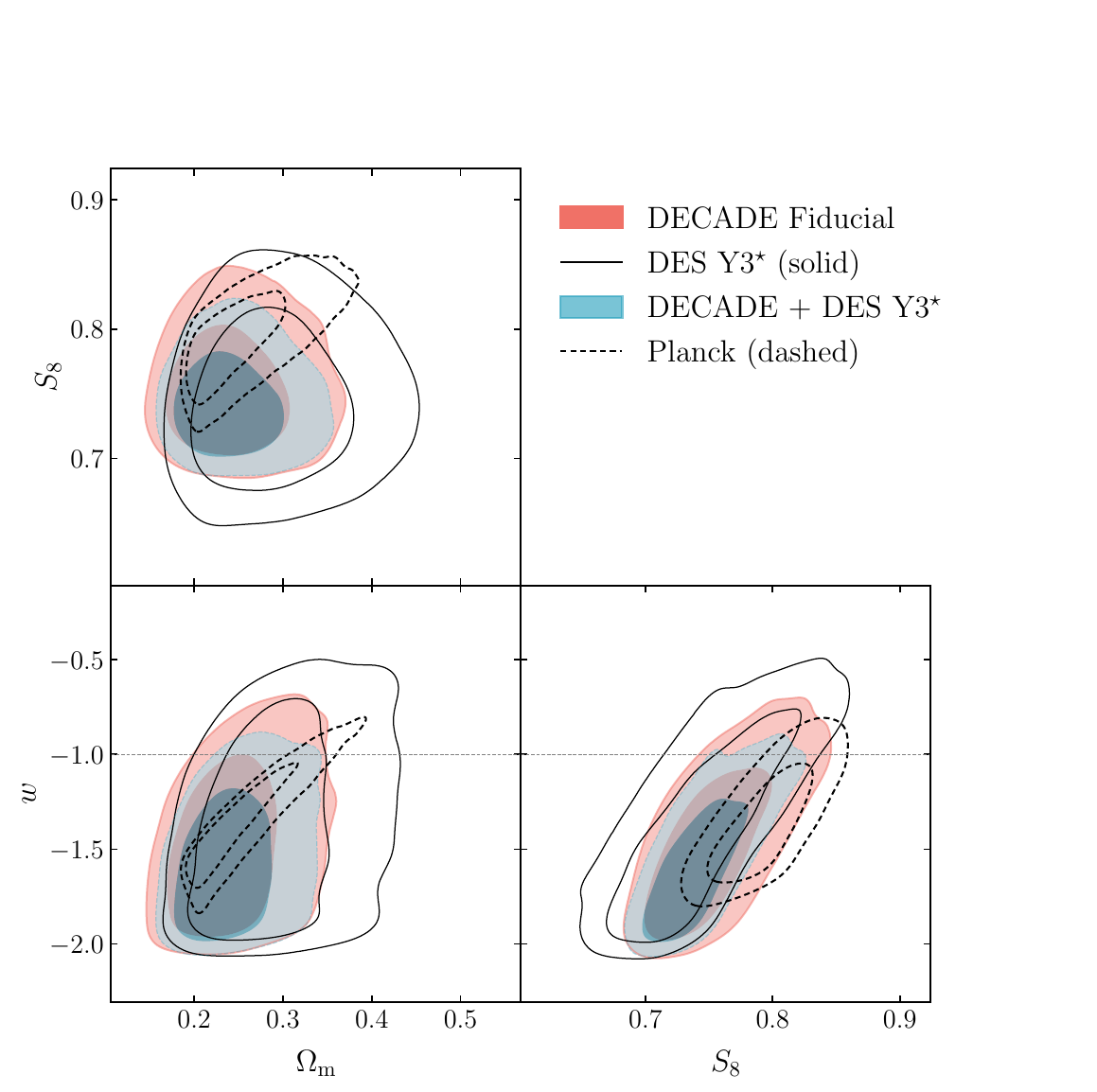}
    \caption{Cosmic shear constraints under a \wCDM model, for $\Omega_{\rm m}$, $S_8$ and $w$ for \decade (red), DES Y3 (black solid), the combination of \decade and DES y3 (blue), and \textit{Planck} (black dashed). The constraint from \decade (\decade + DES Y3) is $w = -1.47^{+0.41}_{-0.25}$ ($w = -1.57^{+0.38}_{-0.16}$), which is consistent with \textit{Planck} and with the \LCDM expectation of $w = -1$ (dotted gray line). The DES Y3 data was reanalyzed using our pipeline. All presented constraints are listed in Table \ref{tab:constraints}.}
    \label{fig:constraints_wcdm}
\end{figure*}

All constraints presented thus far are for a \LCDM model of  the Universe, \ie a model where dark energy has an equation of state $w = -1$. We now promote $w$ to a free parameter of the model. This tests the evolution of the density of dark energy over time and can provide hints on the origin of the accelerated expansion of our Universe \citep[\eg][]{Mortonson:2014:DE, Linder:2023:Benchmarks}. This section presents the constraints for this model, \wCDM.

Figure~\ref{fig:constraints_wcdm} shows the posterior constraints for $\Omega_{\rm m}$, $S_8$ and $w$. Our tests on simulated data vectors (performed post-unblinding) find that the \wCDM constraints on $S_8$ and $w$ incur some mild ($<0.7\sigma$) projection effects on the mean value of the posterior.\footnote{For our \LCDM analysis, we checked the impact of projections effects in \citetalias{paper3} (see their Figure 2) and found it to be negligible in this case.} We stress that such effects do not indicate any bias in the posterior, but simply highlight the nonlinear nature of the posteriors. All qualitative conclusions below, regarding the consistency between different results/constraints, are unchanged even under such effects. We follow the DES Y3 analyses \citep{Secco2021, Amon2021} in using a uniform prior of $w \in [-2, -1/3]$. The \decade constraints under \wCDM are,
\begin{align}
    \Seight & = 0.753^{+0.024}_{-0.041}, \\[4pt]
    \Om & = 0.244^{+0.034}_{-0.057},\\[4pt]
    w & = -1.47^{+0.41}_{-0.25}.
\end{align}
The \wCDM-based constraint on $\Seight$ is shifted $1\sigma$ lower relative to the \LCDM-based constraint, and this is consistent with the behavior found in DES Y3 \citep[][see Line 6 in their Figure 10]{Secco2021}. We note that the DES Y3 cosmic shear analysis does not quote a constraint on $w$ \citep{Secco2021, Amon2021} while the \decade data has a weak constraint on the upper bound of $w$. Regardless, both results are consistent with $w = -1$; with the \decade one being consistent within $1.3\sigma$. The combined result is similarly consistent with $w = -1$ at $1.5\sigma$,\footnote{In both cases, we use the upper-bound uncertainty as the $\sigma$ in our distance metric (Equation~\ref{eq:s8_dist}). This is chosen because the point $w = -1$ is in the upper half of the posteriors.}
\begin{align}
    \Seight & = 0.743^{+0.020}_{-0.032},\\[4pt]
    \Om & = 0.243^{+0.030}_{-0.049},\\[4pt]
    w & = -1.57^{+0.38}_{-0.16}.
\end{align}
The constraints are listed in Table \ref{fig:constraints}. Additionally, Figure \ref{fig:IA_wcdm} shows that the constraints on $w$ are weakly correlated with the TATT IA amplitude $a_2$, and that negative values of $a_2$ can slightly push $w$ to more negative values.

\begin{table*}
    \centering
    \begin{tabular}{cccccccc}
    \hline
    Run & $\Seight$ & $\Om$ & $\sigma_8$ & $w$ & $\chi^2/N_{\rm data}$ & $p$ & FoM$_{\Seight \Om}$\\
    \hline\hline \rule{0pt}{10pt}
    DECADE Fiducial & $0.791^{+0.026}_{-0.032}$ & $0.268^{+0.033}_{-0.050}$ & $0.845^{+0.075}_{-0.092}$ & $-$ & 264.7/220 & 0.021 & 861 \\[3pt]
    DES Y3$^\star$ & $0.779\pm 0.031$ & $0.307^{+0.043}_{-0.071}$ & $0.782^{+0.086}_{-0.10}$ & $-$ & 238.6/222 & 0.212 & 615 \\[3pt]
    DECADE + DES Y3$^{\scalebox{0.7}{$\star$}}$ & $0.791\pm 0.023$ & $0.277^{+0.034}_{-0.046}$ & $0.830^{+0.071}_{-0.082}$ & $-$ & 502.0/442 & 0.025 & 1240 \\[3pt]
    $\textit{Planck}$ 2018 & $0.827\pm 0.018$ & $0.332^{+0.010}_{-0.020}$ & $0.786^{+0.030}_{-0.012}$ & $-$ & --- & --- & 3252 \\[3pt]
    DES Y3$^{\scalebox{0.7}{$\star$}}$, with SR & $0.773\pm 0.026$ & $0.325^{+0.047}_{-0.069}$ & $0.755^{+0.078}_{-0.098}$ & $-$ & 241.0/222 & 0.182 & 847 \\[3pt]
    DECADE + DES Y3$^{\scalebox{0.7}{$\star$}}$, with SR & $0.787\pm 0.020$ & $0.295^{+0.036}_{-0.047}$ & $0.801^{+0.065}_{-0.079}$ & $-$ & 505.4/442 & 0.020 & 1522 \\[3pt]
    \hline \rule{0pt}{10pt}
    DES Y3 & $0.759\pm 0.023$ & $0.290^{+0.041}_{-0.060}$ & $0.783^{+0.075}_{-0.091}$ & $-$ & 239.9/220 & 0.170 & 926 \\[3pt]
    DES Y3, $\Lambda$CDM opt. & $0.772\pm 0.017$ & $0.289^{+0.039}_{-0.054}$ & $0.795\pm 0.073$ & $-$ & 285.7/268 & 0.219 & 1362 \\[3pt]
    KiDS-1000, COSEBIs & $0.751^{+0.024}_{-0.019}$ & $0.286^{+0.056}_{-0.10}$ & $0.79^{+0.12}_{-0.14}$ & $-$ & 82.2/70 & 0.161 & 650 \\[3pt]
    KiDS-1000, $\xi_\pm$ & $0.766\pm 0.018$ & $0.227^{+0.033}_{-0.053}$ & $0.894\pm 0.095$ & $-$ & 152.1/115 & 0.013 & 1165 \\[3pt]
    KiDS-1000, bandpowers & $0.751^{+0.031}_{-0.022}$ & $0.328^{+0.072}_{-0.10}$ & $0.74^{+0.10}_{-0.14}$ & $-$ & 260.3/220 & 0.034 & 588 \\[3pt]
    HSC Y3, $\xi_\pm$ & $0.770\pm 0.030$ & $0.257^{+0.037}_{-0.050}$ & $0.841^{+0.078}_{-0.087}$ & $-$ & 150.0/140 & 0.266 & 786 \\[3pt]
    HSC Y3, $C_{\ell}$ & $0.778\pm 0.030$ & $0.225^{+0.027}_{-0.061}$ & $0.914^{+0.11}_{-0.077}$ & $-$ & 58.5/60 & 0.531 & 681 \\[3pt]
    DES Y3 + KiDS-1000 & $0.790^{+0.018}_{-0.016}$ & $0.280^{+0.037}_{-0.047}$ & $0.825\pm 0.069$ & $-$ & 378.0/348 & 0.129 & 1415 \\[3pt]
    \hline  
    \rule{0pt}{10pt}
    DECADE, $w$CDM & $0.753^{+0.024}_{-0.041}$ & $0.244^{+0.034}_{-0.057}$ & $0.848\pm 0.088$ & $-1.47^{+0.41}_{-0.25}$ & 264.3/220 & 0.022 & 611 \\[3pt]
    DES Y3$^{\scalebox{0.7}{$\star$}}$, $w$CDM & $0.744^{+0.036}_{-0.054}$ & $0.291^{+0.047}_{-0.073}$ & $0.767^{+0.083}_{-0.099}$ & $-1.38^{+0.58}_{-0.23}$ & 234.8/222 & 0.265 & 357 \\[3pt]
    DECADE + Y3$^{\scalebox{0.7}{$\star$}}$, $w$CDM & $0.743^{+0.020}_{-0.032}$ & $0.243^{+0.030}_{-0.049}$ & $0.835\pm 0.077$ & $-1.57^{+0.38}_{-0.16}$ & 504.1/442 & 0.022 & 851 \\[3pt]
    $\textit{Planck}$ 2018, $w$CDM & $0.793^{+0.025}_{-0.028}$ & $0.252^{+0.018}_{-0.056}$ & $0.874^{+0.072}_{-0.042}$ & $-1.37^{+0.26}_{-0.16}$ & --- & --- & 1126 \\[3pt]
    \hline
    \end{tabular}
    \caption{Summary of the constraints from our \LCDM (top) and \wCDM (bottom) analyses compared against other lensing results (middle). The constraints are shown in Figure \ref{fig:constraints} and \ref{fig:constraints_wcdm}. From left to right we list the constraints on cosmological parameters ($S_8, \Om, \sigma_8, w$), the reduced $\chi^2$ of the best-fit cosmology, the associated $p$-value of that best fit, and the Figure-of-Merit $\text{FoM}_{\Seight\Om} = \det[\text{Cov}(\Seight, \Om)]^{-1/2}$. We use ``DES Y3$^{\scalebox{0.7}{$\star$}}$'' to denote the reanalysis of DES data using our pipeline (see Appendix \ref{appx:DES_matched}). All \textit{Planck} constraints come from the ``TT+TE+EE+LowE'' dataset and use our priors for the cosmological parameters (see Section \ref{sec:lcdm} for more details). We also compare against public results from DES Y3 \citep{Secco2021, Amon2021}, KiDS-1000 \citep{Asgari2021}, HSC Y3 \citep{Li2023, Dalal2023} and the DES Y3 and KiDS-1000 joint analysis \citep{DESKiDS2023}. Our constraints on all cosmological parameters are in good agreement with other surveys, for both \LCDM and \wCDM models. Amongst the public DES constraints shown in the middle rows, the combined analysis of DES Y3 and KiDS-1000 matches our choice in not using the DES shear ratio measurements \citep{Sanchez2022}, whereas the DES-only constraints do use this information.}
    \label{tab:constraints}
\end{table*}

\section{Discussion}\label{sec:Discussion}

\begin{figure*}
    \centering
    \includegraphics[width = 1.8\columnwidth]{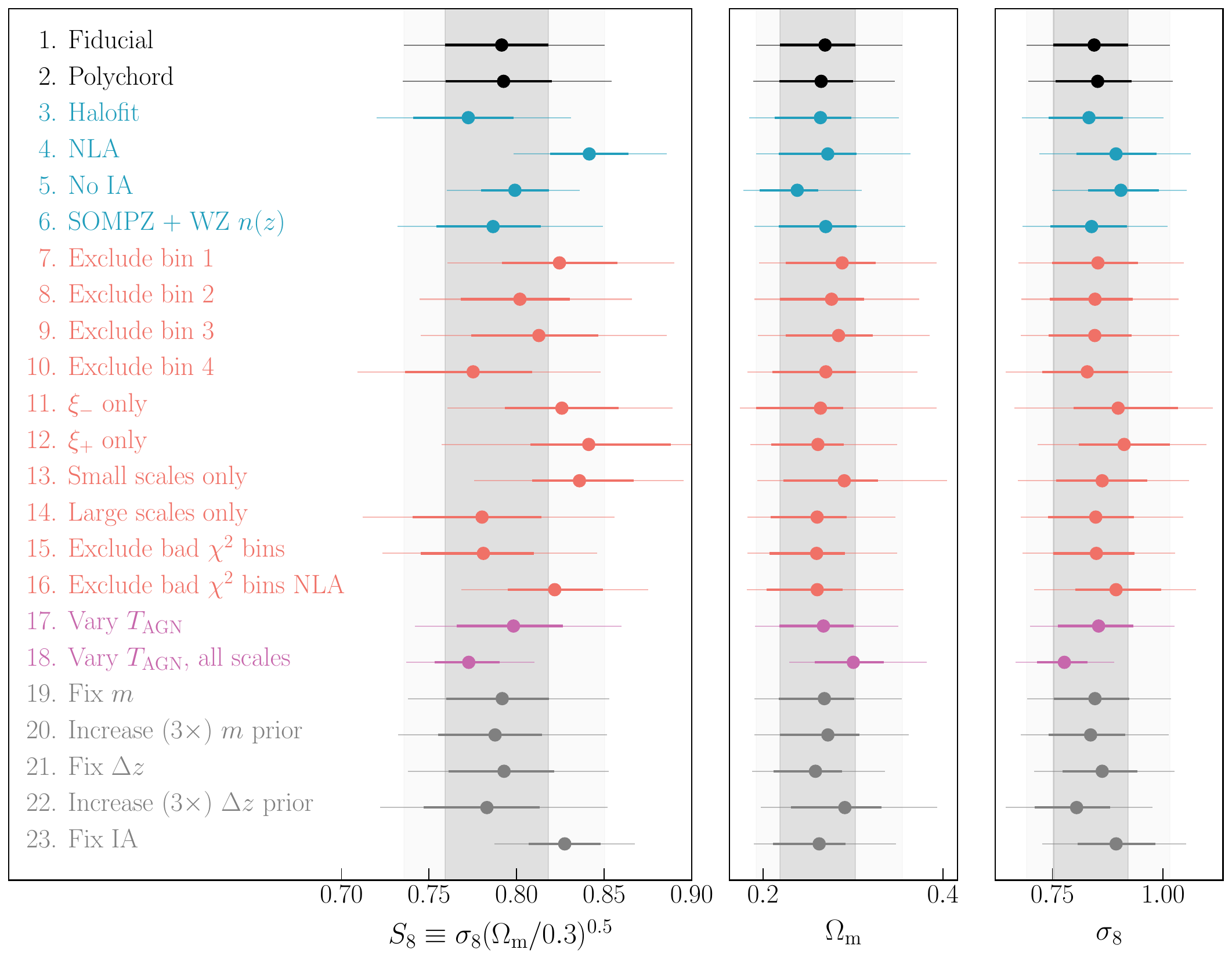}
    \caption{\decade constraints in $S_8$, $\Omega_{\rm m}$, and $\sigma_8$ for different variant analyses under our fiducial \LCDM pipeline. The fiducial result is shown in Line 1, and as a vertical gray band (denoting the $1\sigma$ and $2\sigma$ interval) to facilitate comparison with other results. Line 2 shows our results are consistent across different samplers. Lines 3--6 (blue) considers alternative model assumptions. Lines 7--16 (red) show constraints when only using part of our data vector. Lines 17--18 (purple) investigates changes to our result when we incorporate the small scales (which were previously discarded) and model baryonic corrections via \textsc{HMCode} \citep{Mead2021b} using an additional parameter ($T_{\rm AGN}$). Lines 19--23 (gray) showcase the sensitivity of our constraints to our priors on the nuisance parameters. The detailed discussion of these points can be found in Section~\ref{sec:Discussion}, while the numerical constraints are listed in Table \ref{tab:data_variation}. All results are obtained after using scales cuts. See Section \ref{sec:sec:dropdata} for more details on the analysis and the discussion of results.}
    \label{fig:data_variation}
\end{figure*}

To demonstrate the robustness of our cosmological constraints, we perform a number of additional tests on our analysis results. We summarize these in Figure~\ref{fig:data_variation} and Table \ref{tab:data_variation}. We discuss below our interpretation of all these results. A subset of the tests mimic those we performed in \citetalias{paper3} with simulated data vectors. A few of them are also connected to our pre-unblinding analysis and unblinding criteria, both of which are described in detail in Appendix \ref{appx:unblind_test}.

\subsection{Goodness of fit} \label{sec:sec:gof}

The two best-fit models shown in Figure \ref{fig:datavec} have a $\chi^2 \approx 264$ which, for $N_{\rm data} = 220$, has a $p$-value of $0.021$, corresponding to a $\sim 2\sigma$ discrepancy between data and model (see Table \ref{tab:constraints}). While this passes our unblinding criteria (see Appendix~\ref{appx:unblind_test}), the $p$-value is still somewhat on the lower side and motivated us to examine the fit in more depth.\footnote{We also note that our $p$-value passes the criteria set by other surveys \citep{Secco2021, Amon2021, Asgari2021} which varies between the thresholds $p > 0.001$ and $p > 0.01$.} We isolate the primary source of the high $\chi^2$ to $\xi_+$ measurements in three tomographic bin pairs: $(1,1)$, $(2,4)$, $(4,4)$. There are no unique features that distinguish these bins from the rest, \eg they are not all on the higher redshift end of the sample \etc Discarding these bin pairs brings the $p$-value to under $1\sigma$, but does not change our final cosmology constraint, \ie the measurements contributing to the relatively high $\chi^2$ of the total data vector do not push the cosmology constraints in a specific direction. The ``oddity'' in these bin pairs is their somewhat large, seemingly uncorrelated, scatter. To see if this large scatter correlates with image quality (or \textsc{Metacalibration} quantities), we visually inspected data vectors from the forty-six subsets of the catalog analyzed in \citetalias{paper3} but did not find any particular split that correlated with this larger scatter. Furthermore, when considering the goodness-of-fit for each of these forty-six splits, almost all splits have a $\chi^2$ notably better than that found in our Fiducial case with the remaining splits exhibiting a $\chi^2$ that is still similar or slightly better than that of the Fiducial case. Given these findings, we conclude that the origin of the scatter and the high $\chi^2$ values is indeed from statistical fluctuations.

% \begin{itemize}
    % \item reference to appendix that we know the gof is slightly low, but did extensive test to make sure there is no known things that we are doing wrong
    % \item discuss that the 3 bins are contributing to most of the bad chi2
    % \item removing them does not change the constraint or fix the IA thing discussed below -- they could indeed just be statistical fluctuations
% \end{itemize}

\begin{table}[]
    \centering
    \begin{tabular}{|cccc|}
    \hline
    Run & $S_8$ & $\Omega_{\rm m}$ & $\chi^2/N_{\rm data}$ ($p$)\\[2pt]
    \hline \rule{0pt}{10pt}
    Fiducial & $ 0.791^{+0.027}_{-0.032}$ & $ 0.269^{+0.034}_{-0.050}$ & 1.20 (0.021) \\[3pt]
    Polychord & $ 0.793^{+0.028}_{-0.033}$ & $ 0.265^{+0.035}_{-0.046}$ & 1.21 (0.019) \\[3pt]
    Halofit & $ 0.772^{+0.026}_{-0.031}$ & $ 0.264^{+0.034}_{-0.051}$ & 1.21 (0.019) \\[3pt]
    NLA & $ 0.841\pm 0.022$ & $ 0.272^{+0.032}_{-0.054}$ & 1.29 (0.003) \\[3pt]
    No IA & $ 0.799\pm 0.019$ & $ 0.238^{+0.023}_{-0.042}$ & 1.32 (0.001) \\[3pt]
    SOMPZ + WZ $n(z)$ & $ 0.787^{+0.027}_{-0.032}$ & $ 0.270^{+0.034}_{-0.052}$ & 1.21 (0.016) \\[3pt]
    \hline \rule{0pt}{10pt}
    Exclude bin 1 & $ 0.824\pm 0.034$ & $ 0.288^{+0.037}_{-0.062}$ & 1.21 (0.051) \\[3pt]
    Exclude bin 2 & $ 0.802^{+0.029}_{-0.034}$ & $ 0.276^{+0.036}_{-0.058}$ & 1.22 (0.046) \\[3pt]
    Exclude bin 3 & $ 0.813^{+0.034}_{-0.039}$ & $ 0.284^{+0.038}_{-0.059}$ & 1.28 (0.016) \\[3pt]
    Exclude bin 4 & $ 0.775^{+0.034}_{-0.039}$ & $ 0.270^{+0.034}_{-0.059}$ & 1.19 (0.066) \\[3pt]
    $\xi_-$ only & $ 0.826\pm 0.033$ & $ 0.264^{+0.025}_{-0.072}$ & 1.14 (0.203) \\[3pt]
    $\xi_+$ only & $ 0.841^{+0.047}_{-0.033}$ & $ 0.261^{+0.028}_{-0.052}$ & 1.18 (0.060) \\[3pt]
    Small scales only & $ 0.836^{+0.031}_{-0.027}$ & $ 0.290^{+0.037}_{-0.067}$ & 1.18 (0.090) \\[3pt]
    Large scales only & $ 0.780^{+0.034}_{-0.040}$ & $ 0.260^{+0.033}_{-0.051}$ & 1.10 (0.228) \\[3pt]
    Exclude bad $\chi^2$ bins & $ 0.781^{+0.029}_{-0.036}$ & $ 0.260^{+0.032}_{-0.052}$ & 0.98 (0.566) \\[3pt]
    Exclude bad $\chi^2$ bins NLA & $ 0.822\pm 0.027$ & $ 0.260^{+0.028}_{-0.056}$ & 1.06 (0.290) \\[3pt]
    \hline \rule{0pt}{10pt}
    Vary $T_{\rm AGN}$ & $ 0.798^{+0.028}_{-0.032}$ & $ 0.267^{+0.033}_{-0.049}$ & 1.19 (0.025) \\[3pt]
    Vary $T_{\rm AGN}$, all scales & $ 0.773^{+0.018}_{-0.020}$ & $ 0.301^{+0.034}_{-0.043}$ & 1.14 (0.031) \\[3pt]
    \hline \rule{0pt}{10pt}
    Fix $m$ & $ 0.792^{+0.027}_{-0.032}$ & $ 0.268^{+0.033}_{-0.051}$ & 1.20 (0.022) \\[3pt]
    Increase $(3\times)$ $m$ prior & $ 0.788^{+0.027}_{-0.033}$ & $ 0.272^{+0.035}_{-0.053}$ & 1.20 (0.022) \\[3pt]
    Fix $\Delta z$ & $ 0.793\pm 0.030$ & $ 0.258^{+0.029}_{-0.047}$ & 1.23 (0.012) \\[3pt]
    Increase $(3\times)$ $\Delta z$ prior & $ 0.783^{+0.030}_{-0.036}$ & $ 0.291^{+0.040}_{-0.060}$ & 1.19 (0.026) \\[3pt]
    Fix IA & $ 0.827\pm 0.021$ & $ 0.263^{+0.029}_{-0.051}$ & 1.30 (0.002) \\[3pt]
    \hline
    \end{tabular}
    \caption{The results from the various runs shown in Figure \ref{fig:data_variation}. From left to right, we list the run name, the constraints on $S_8$ and $\Om$, and the reduced $\chi^2$ along with the associated $p$-value given in parentheses. 
    }
    \label{tab:data_variation}
\end{table}

\subsection{Intrinsic alignments} \label{sec:sec:IA}

During the pre-unblinding analysis, we decided to use TATT as our fiducial IA model since using NLA resulted in a noticeably poorer goodness-of-fit (see Table \ref{tab:data_variation} and Section \ref{appx:unblind_test}). Furthermore, TATT reduces to the NLA model when fixing three of the five free parameters ($a_2, b_{\rm TA}, \eta_2 = 0$). After unblinding, we computed the Bayesian evidence ratios from the posteriors \citep[\eg][see their Equation 7]{Secco2021} and find a clear preference in our data for TATT over NLA and over the ``no IA'' model. In particular NLA (no IA) is disfavored with a Bayesian evidence ratio of 0.013 (0.006) relative to the TATT model. This is consistent with our statement above that the goodness-of-fit improves considerably when using TATT over NLA. We note that the evidence ratios continue to favor TATT over NLA, with an evidence ratio of 0.042 for the latter, even if we exclude the three bin pairs that contribute most to degrading the goodness-of-fit (see Section \ref{sec:sec:gof} above). Thus, our data has a clear preference for the TATT model.

Our fiducial TATT constraints show the \decade data prefer a non-zero value for $a_1$ ($a_2$) at $2.8\sigma$ ($2.9\sigma$),
\begin{align}
    a_1 & = 0.73^{+0.22}_{-0.46}\\[4pt]
    a_2 & = -2.47^{+0.56}_{-0.65}.
\end{align}
There is no well-motivated prior expectation for the values of IA amplitudes (hence our wide prior) since the values depend on the exact galaxy sample being selected. However, our constraints on $a_1$ and $a_2$ are statistically consistent with the (broad) posteriors from DES Y3 \citep{Secco2021, Amon2021} at $0.9\sigma$ and $1.5\sigma$, respectively. Figure \ref{fig:IA} shows this comparison in more detail. The redshift evolution coefficients, $\eta_{1, 2}$, are not well constrainted but we discuss their behavior in detail in Appendix \ref{appx:IA}. The bias parameter $b_{\rm TA}$ is also unconstrained, consistent with the DES Y3 results, and we do not discuss this further.

The \decade results find a statistically significant preference for non-zero IA amplitudes that is qualitatively consistent with other works \citep{Samuroff:2019:IA, Asgari2021, DESKiDS2023}; though, it is in contrast to the DES Y3 cosmic shear results \citep{Secco2021, Amon2021}, which find no preference for an IA signal. For the source galaxy sample from DES Y3, \citet{McCullough:2024:Blueshear} find that roughly 60\% is comprised of blue galaxies while the remaining are a combination of red galaxies and galaxies with ambiguous classifications. We expect the DECADE source galaxy sample to exhibit similar characteristics given its similarity to the DES Y3 sample.

Blue galaxies are expected to have a weaker IA signal relative to their red counterparts \citep[\eg][]{Blazek2019}; this is corroborated by direct measurements of the IA signal \citep[\eg][]{Samuroff:2019:IA, Samuroff2022, Georgiou:2025:IA, Siegel:2025:IA} though the uncertainties---particularly for amplitude of IA in blue galaxies---are still larger than those found in our analysis. Our source galaxy sample, without any color splits, could still exhibit detectable IA signals depending on the nature the red galaxies in the sample. A consistent mapping of direct IA constraints to cosmic shear results is an ongoing effort in the community. In summary, it is possible that our constraint above reflects (at least partly) a real IA signal in the data. However, to be conservative, we posit other reasons our constraints may prefer non-zero values.

The IA parameter constraints can shift due to noisy data: \citet[][see their Figure 15]{Amon2021} show that noisy, simulated data vectors can generate posteriors that explore large values for $a_1, a_2$. We have also verified this result through our own simulation tests. The IA parameters have their own redshift scaling ($\eta_{1, 2}$) which are free parameters. Therefore the IA model can also compensate for inaccurate estimates of the $n(z)$ distribution (\ie if the priors on the redshift calibrations are too narrow) by scaling the total correlation (weak lensing plus IA) in a given redshift bin relative to the others \citep{Leonard:2024:IARedshifts}. Figure \ref{fig:data_variation} shows a crude check of this effect, where broadening the redshift uncertainty by $3\times$ gives consistent constraints on $S_8$. We have verified the IA constraints are similarly consistent if we widen this prior.

To probe this further, we extract the contributions from the individual IA terms and show them in Figure \ref{fig:IA_contrib}. That figure, and the discussion accompanying it, finds that the preference for non-zero IA parameters originates from a slightly altered scale-dependence seen in the $\xi_+$ measurements (and somewhat from the $\xi_-$ measurements as well), relative to a lensing-only (no IA) prediction. This trend is found in measurements across all bin pairs and changes mildly with redshift. We inspected our data vector and IA constraints across the forty-six catalog-level splits in \citetalias{paper3}, where the splits are defined on image quality and object properties, and do not find any observational quantities that cause the preference for non-zero $a_1, a_2$ values. In fact, almost all of the forty-six splits also show a preference for $a_1 > 0$ and $a_2 < 0$, consistent with the Fiducial case. So, while our data prefers $a_1 > 0$ and $a_2 < 0$, the tests from \citetalias{paper3} (see their Table B1 and B2) find that this preference is not driven by a specific subset of the galaxies (\eg only those observed in particularly good/poor seeing or only those with small/large angular sizes). We also note that \citetalias{paper1} (see their Figure 13) finds that our measured galaxy shears exhibit no correlation with the spatial distribution of image quality. This indicates the measured shears have no statistically significant contamination from the variation in image quality, which in turn implies that such image quality variations do not impact the IA constraints (as the IA constraints are derived from the measured shears).

Some of the data variant runs in Figure~\ref{fig:data_variation} also show changes in their IA posteriors (see Figure \ref{appx:IA}); we discuss this further in Section \ref{sec:sec:dropdata}. Specifically, they exhibit shifts to $a_2 \approx 0$. The $a_1$ posterior is relatively similar across the data vector-level splits shown in that figure and then $\eta_2$, while still pushing against the upper bound of the prior, has a broader posterior that is now consistent with $\eta_2 = 0$. Removing subsets of the data vector causes a weakening in the previously mentioned trends with redshift and angular scales. This subsequently changes the IA posteriors. We explicitly show in Section \ref{sec:sec:dropdata} below that the $S_8$ constraints from these different variant analysis are all statistically consistent with each other. We direct the interested reader to Appendix \ref{appx:IA} for a more extensive discussion on our IA results.

At this time, we do not have a clean explanation for the exact origin of the IA constraint in our data. Our empirical tests (discussed above) indicate the underlying signal does not come from variations in image quality, nor from a specific subset of the data (\eg galaxies with large angular sizes). We are unable to pursue deeper interpretations of this signal using weak lensing measurements alone. Direct measurements of IA using cross-correlations with spectroscopic data will be able to shed more insight into the signal \citep[\eg][]{Samuroff:2019:IA, Samuroff2022, Georgiou:2025:IA}. However, we emphasize that we have confirmed (via the discussion in this Section, and also in Appendix \ref{appx:IA}) that our final cosmology constraints are insensitive to any oddities in the IA constraints.

\subsection{Constraints from subsets of the data vector: redshift bins, $\xi_+$ vs. $\xi_-$, large scale vs. small scale} \label{sec:sec:dropdata}

An important validation of our final results is checking that different subsets of the data vector provide consistent constraints --- \ie checking that the different subsets of the data vector are internally consistent with each other. We do this by excluding different parts of the data vector shown in Figure \ref{fig:datavec} and reanalyzing our parameter constraints. Our tests are in Lines 7--16 (red) in Figure \ref{fig:data_variation}. We consider four such tests:
\begin{itemize}
    \item exclude one tomographic bin at a time, including all cross-correlations of that bin with others,\footnote{We also redid this analysis excluding the two lowest/highest redshift bins at once. However, in practice, this discards a majority of the data-vector and the resulting constraints on $S_8$ are consistent as the posteriors are significantly broader. The result of the test is therefore uninformative.}
    \item use only $\xi_+$ or $\xi_-$,
    \item split the data vector (post scale-cuts) to use half of the smallest/largest scales in each bin pair,
    \item exclude three bin pairs that contribute most to our relatively high $\chi^2$ (see Section \ref{sec:sec:gof}).
\end{itemize}

Excluding different tomographic bins changes the constraints by less than $1\sigma$, as does excluding the three bin pairs that predominantly increase the $\chi^2$ between the measurements and the best-fit model. Using only $\xi_+$ or $\xi_-$ similarly changes the constraints by $\lesssim 1 \sigma$.\footnote{It may be surprising that the $\xi_-$ constraining power is fairly similar to that of $\xi_+$, even though the former has half as many datapoints after scale cuts. This is because both constraints are limited by uncertainties in the IA model, and the latter uncertainty is the same/similar across both analyses, resulting in similar posterior widths on $S_8$. Similar behavior is found in the DES Y3 cosmic shear analysis \citep[][see their Figure 7]{Amon2021}} Finally, using only small scales (large scales) shifts the constraint by $1.07\sigma$ ($0.28\sigma$). Figure \ref{fig:IA} shows the IA posteriors for a few variant analyses. The posteriors are broadly consistent with each other, though the constraint on $a_2$ is now broader and consistent with $a_2 = 0$.

We also note that three of the points --- ``$\xi_+$ only'', ``$\xi_-$ only'', and ``small scales only'') all shift to higher $S_8$. Given we only have three points in this set, and all shifts are approximately within $1\sigma$, the results are all still statistically consistent. Furthermore, analysing noisy data with IA may cause seemingly correlated shifts. To explicitly test the significance of these shifts, we perform simulated tests of the exact analysis configuration, where we (i) generate a theoretical data vector at the best-fit value of the cosmology and nuisance parameters; (ii) add noise using the analytic covariance model, and; (iii) reanalyze the data vector under the ``$\xi_+$/$\xi_-$ only'' and ``large/small scale only'' analysis configuration. We run the four variant analyses on thirty noisy data vectors and extract $S_8$ in each case. We estimate the probability that our observed shifts in the four variant, ``exclude data'' constraints are consistent with noise, and find $p \!\!\sim\!\! 0.1$. We obtain this by fitting a 4D multivariate Gaussian to the thirty best-fit values per analysis configuration, and finding the probability-to-exceed for the observed shifts. Other analysis choices --- such as using kernel density estimators, incorporating the full posterior, \etc --- give probabilities that vary by $\Delta p \approx 0.05$. In the end, this is only a crude estimate of probability but shows that it is indeed possible for noise to generate the observed shifts.\footnote{We have also redone this exercise using a simulated data vector with the same best-fit cosmology but now removing the IA signal ($a_1 = a_2 = 0$). In this case, we find $p = 0.08$, which suggests that our probability estimate above is not that sensitive to our choice for the IA signal used in the simulated data vector.}

While we do not find any evidence that these shifts in the three points are systematics-driven rather than statistical fluctuations, we still briefly consider a thought experiment where the entire shift is due to systematics alone. Figure \ref{fig:data_variation} shows the shift is at most $\lesssim 1\sigma$. Thus, even if the observed shifts in $\Seight$ are completely systematics-driven, the shifts are still within the posterior of the Fiducial constraints. This last point is simply a restatement of our discussion above, that the three constraints are statistically consistent with the Fiducial one.

\subsection{Small-scale modeling} \label{sec:sec:Baryons}

As discussed above, and also in \citetalias{paper3}, a current limitation in weak-lensing analyses is uncertainty in the model for the small-scale power spectrum --- particularly, for the signal of baryons on these scales. In our fiducial setup, we use the approach of DES Y3 \citep{Secco2021, Amon2021} and remove scales where baryon corrections have a statistically noticeable impact on our data vector. In this section, we follow \citet{DESKiDS2023} in performing a variant analysis where we (i) use the same scale cuts as the Fiducial analysis but also marginalize over baryon corrections, and (ii) marginalize over baryon corrections but using all available datapoints, without imposing any scale cuts. We also follow that \citet{DESKiDS2023} in including the baryon corrections using the model of \citet{Mead2021b}. The corrections are parameterized by a single amplitude, $T_{\rm AGN}$, which can be interpreted as an ``effective feedback'' parameter --- larger values indicate stronger suppression of matter clustering. We use a prior of $T_{\rm AGN} \in [10^{7.6}, 10^{8.0}]$, which is the range the model was calibrated to and is the same prior used in other works \citep{DESKiDS2023, Bigwood:2024:BaryonsWLkSZ}.

Figure \ref{fig:data_variation} shows the results for the marginal $\Seight$ constraints while the posterior on other parameters are shown in Figure \ref{fig:Tagn}. In general, the constraints from these two variants are well within $1\sigma$ of our Fiducial constraint. The inclusion of a baryon component to our model does not change the goodness-of-fit by any appreciable amount, and does not change the IA constraints either. The latter is consistent with simulation measurements that also find a weak connection between IA and baryons on the scales relevant to our analysis \citep[\eg][]{Tenneti:2017:BaryonIA, Soussana:2020:BaryonIA}. The shifts in $\Seight$, relative to the Fiducial constraints, are $0.15\sigma$ and $0.54\sigma$ when using/ignoring scale cuts, respectively. 

% \begin{itemize}
%     \item Tagn discussion, seems like we do get more info from small scales -- a big improvement
%     \item should look at gof and IA, does it change anything?
% \end{itemize}

\subsection{Nuisance parameters} \label{sec:sec:Nuisance}

In Lines 19--23, we either reduce or increase the priors on the nuisance parameters (the multiplicative bias and mean redshift uncertainty for each of the four tomographic redshift bins) and investigate the impact on the cosmological constraints. We note that the nuisance priors ($m$, $\Delta z$) are derived from \citetalias{paper1} and \citetalias{paper2}, respectively, and have been validated extensively. When changing the priors on these parameters to a delta function (line 19 and 21), we are checking how much a given nuisance parameter contributes to the final posterior width on $S_8$. We find that for both $m$ and $\Delta z$ there is nearly no change in the $S_8$ constraints; this means our current uncertainties on $m$ and $\Delta z$ are not the limiting factors in the analysis. On the other hand, by increasing the prior width by 3$\times$, we check the impact on cosmology if we incorrectly assumed overly tight priors. We find that for $m$ the impact is still negligible. This is not surprising as our priors on $m$ are quite small (relative to DES Y3) due to the larger number of image simulations we used \citepalias{paper1}. On the other hand, increasing the $\Delta z$ prior results in a degradation in $S_8$ constraining power by $\sim\!\!10\%$. We also note that the mean value of $S_8$ shifts negligibly in the latter case. We note the constraining power on $\Om$ is degraded slightly more, by $\sim 15\%$.

We now move onto the IA parameters. If we fix the IA priors so that all the TATT parameters are delta functions, we find that the constraints on $S_8$ improve by nearly 30\%. This makes explicit our previous statement that the final constraints are limited by uncertainties in the IA signals. The only difference between our ``No IA'' and ''Fix IA'' results is the former sets all IA amplitudes to zero whereas the latter sets them to the Fiducial values found in the DES 3$\times$2-point\footnote{The 3$\times$2 nomenclature of DES refers to the combination of two-point correlation functions of three probes: galaxy clustering, galaxy galaxy-lensing, and cosmic shear.} analysis \citep{DESY3KP2022}; see Table 1 in \citetalias{paper3} for the precise values. Both results show that the IA modeling is the limiting uncertainty in our analysis.

\section{Summary}
\label{sec:summary}

% \begin{itemize}
% \item summarize the whole decade cosmic shear program again
% \item our fiducial constraints, tensions
% \item decade + des constraints, tensions
% \item we did a bunch of things to test the data and the results are robust
% \item significance and implications of cosmology from a dataset like decade, and implications for future surveys
% \end{itemize}

%As mentioned in our introduction, t
In this paper, we present the cosmology results of the \decade cosmic shear project. This work is the fourth in a series of papers detailing the rigorous testing, validation, and calibrations of the various data and modeling pipelines used for this project \citepalias{paper1, paper2, paper3}. The \decade data is an amalgamation of over ten years of community-led observing with the Dark Energy Camera (DECam). The general-purpose catalog will be presented as part of Data Release 3 from the DECam Local Volume Exploration survey \citep[DELVE,][]{Drlica-Wagner:2021, Drlica-Wagner:2022}.

We have carried out a cosmic shear analysis with the \decade data, closely following the methodologies of the DES Y3 cosmic shear pipeline \citep[\eg][]{y3-shapecatalog, Myles:2021:DESY3, Secco2021, Amon2021}. Our catalog contains $107$ million galaxies across $5,\!412 \deg^2$ of the sky spanning the northern Galactic cap and is completely independent of the DES Y3 footprint. The data have a slightly lower source galaxy number density relative to DES Y3 but covers a slightly larger area, and so have similar constraining power. Our fiducial cosmology constraint, under the \LCDM model, is $\Seight = 0.791^{+0.026}_{-0.032}$ and $\Om = 0.268^{+0.033}_{-0.050}$. The constraining power is similar to DES Y3 when both \decade and DES Y3 are analyzed with the same pipeline.

Because the \decade constraints are consistent with those of DES Y3, we are able to combine the two datasets. Furthermore, given they cover independent patches of the sky, we can perform the combination at the likelihood level. The combined cosmic shear analysis spans $\approx\!\! 10,\!000 \deg^2$, and gives \LCDM constraints of $\Seight = 0.791\pm 0.023$ and $\Om = 0.277^{+0.034}_{-0.046}$. All constraints are consistent with previous lensing surveys as well as with the \textit{Planck} 2018 constraints \citep{Planck:2020:Cosmo} from the CMB. When extending the model to \wCDM, the \decade data yield $w = -1.47^{+0.41}_{-0.25}$ while the combination of \decade and DES Y3 gives $w = -1.57^{+0.38}_{-0.16}$. These are likewise consistent with constraints from \textit{Planck} 2018 and with the \LCDM expectation of $w = -1$. The \decade constraints (and the \decade and DES Y3 combined constraints) are consistent with other lensing and CMB surveys to within $1\sigma$ to $1.5\sigma$. All results, and their associated comparisons, are summarized in Table \ref{tab:constraints}.

We have performed extensive tests on the internal consistency of the data vector, as well as of the constraints' sensitivity to different modeling choices, and verified our constraints are robust. These are discussed in detail in Section \ref{sec:Discussion} as well as the relevant appendices referred to therein. We also highlight our ``split tests'' in \citetalias{paper3}, which involved splitting the dataset into subsets ---  based on sky location or some chosen object property --- and rerunning the entire end-to-end cosmic shear analysis pipeline (shear calibration, redshift estimation, covariance modeling, \etc). We used forty-six different splits on catalog and survey properties and found that all resulting constraints are consistent with the Fiducial constraint within $2\sigma$. This is the first time these tests have been pursued at such scale in a cosmic shear analysis, and our work shows their utility in characterizing a wide range of systematics in lensing datasets.

As mentioned previously, the \decade dataset is an amalgamation of many different community programs. This results in significantly larger variations in image quality (seeing, exposure time, depth, \etc) relative to other modern lensing surveys. These variations occur at the individual image-level and also across the sky. The \decade project shows that a cosmic shear analysis with such a catalog, using established techniques from DES Y3, is still able to deliver robust constraints on $S_8$ at the $2-3\%$ level. This work provides a reference point on the usability of imaging data that falls short of the ``ideal'' requirements for a weak lensing survey. Our results encourage exploring the construction of weak lensing datasets using imaging data from wider ranges of image quality/depth. Relaxing such criteria can significantly improve the size of the lensing dataset from a given survey, and therefore the precision of the survey as a whole. A salient, related point is that we find our constraints are limited dominantly by uncertainties in the IA modeling. This is similar to the findings from DES Y3 and further highlights the existing need, even in Stage-III surveys, for a better characterization of the IA signal.

Finally, we note that there is available DECam (community-derived) data in the \textit{southern} Galactic cap, \textit{surrounding} the DES footprint. These imaging data have been processed by \decade, and the associated lensing catalogs, redshift distribution, calibrations, \etc have now been generated (using the same pipelines developed and discussed in this series of papers). The cosmology constraints from this extended \decade dataset are presented in \href{\#cite.paper5}{Anbajagane \& Chang et al. (\citeyear{paper5})}. The full \decade survey fills in much of the sky not covered by DES. Together, the two surveys provide a cosmology analysis spanning $13,\!000\deg^2$, with coverage across most of the sky below ${\rm Dec.} \lesssim 30\degr$. This dataset provides the community with an early preview of a Rubin LSST-esque dataset (though at shallower depth), which can be used to test techniques being developed for the Rubin LSST wide-field data and to also complement the many other experiments observing in this region of the sky that rely on optical/lensing data for their analyses/calibrations.

As we enter the next-generation of gravitational lensing experiments, the \decade cosmic shear project showcases the full power and sophistication of one of the leading current-generation instruments, DECam. The continual use of DECam by the astronomical community and significant investments in its data processing pipelines have made DECam a vital tool in a wide variety of astronomical studies. Nearly two decades after it was first conceived, DECam has now imaged nearly three-quarters of the sky in multiple bands, and the combination of the \decade and DES processing campaigns provide high-fidelity cosmic shear catalogs $>10,\!000\deg^2$ of the sky. This will serve as a useful legacy dataset as the next stage of cosmic shear experiments commence.

\section*{Acknowledgements}

DA is supported by the National Science Foundation (NSF) Graduate Research Fellowship under Grant No.\ DGE 1746045. 
CC is supported by the Henry Luce Foundation and Department of Energy (DOE) grant DE-SC0021949. 
The DECADE project is supported by NSF AST-2108168 and AST-2108169.
The DELVE Survey gratefully acknowledges support from Fermilab LDRD (L2019.011), the NASA {\it Fermi} Guest Investigator Program Cycle 9 (No.\ 91201), and the NSF (AST-2108168, AST-2108169, AST-2307126,  AST-2407526, AST-2407527, AST-2407528). This work was completed in part with resources provided by the University of Chicago’s Research Computing Center. The project that gave rise to these results received the support of a fellowship from "la Caixa" Foundation (ID 100010434). The fellowship code is LCF/BQ/PI23/11970028. C.E.M.-V. is supported by the international Gemini Observatory, a program of NSF NOIRLab, which is managed by the Association of Universities for Research in Astronomy (AURA) under a cooperative agreement with the U.S. National Science Foundation, on behalf of the Gemini partnership of Argentina, Brazil, Canada, Chile, the Republic of Korea, and the United States of America.

Funding for the DES Projects has been provided by the U.S. Department of Energy, the U.S. National Science Foundation, the Ministry of Science and Education of Spain, 
the Science and Technology Facilities Council of the United Kingdom, the Higher Education Funding Council for England, the National Center for Supercomputing 
Applications at the University of Illinois at Urbana-Champaign, the Kavli Institute of Cosmological Physics at the University of Chicago, 
the Center for Cosmology and Astro-Particle Physics at the Ohio State University,
the Mitchell Institute for Fundamental Physics and Astronomy at Texas A\&M University, Financiadora de Estudos e Projetos, 
Funda{\c c}{\~a}o Carlos Chagas Filho de Amparo {\`a} Pesquisa do Estado do Rio de Janeiro, Conselho Nacional de Desenvolvimento Cient{\'i}fico e Tecnol{\'o}gico and 
the Minist{\'e}rio da Ci{\^e}ncia, Tecnologia e Inova{\c c}{\~a}o, the Deutsche Forschungsgemeinschaft and the Collaborating Institutions in the Dark Energy Survey. 

The Collaborating Institutions are Argonne National Laboratory, the University of California at Santa Cruz, the University of Cambridge, Centro de Investigaciones Energ{\'e}ticas, 
Medioambientales y Tecnol{\'o}gicas-Madrid, the University of Chicago, University College London, the DES-Brazil Consortium, the University of Edinburgh, 
the Eidgen{\"o}ssische Technische Hochschule (ETH) Z{\"u}rich, 
Fermi National Accelerator Laboratory, the University of Illinois at Urbana-Champaign, the Institut de Ci{\`e}ncies de l'Espai (IEEC/CSIC), 
the Institut de F{\'i}sica d'Altes Energies, Lawrence Berkeley National Laboratory, the Ludwig-Maximilians Universit{\"a}t M{\"u}nchen and the associated Excellence Cluster Universe, 
the University of Michigan, NSF's NOIRLab, the University of Nottingham, The Ohio State University, the University of Pennsylvania, the University of Portsmouth, 
SLAC National Accelerator Laboratory, Stanford University, the University of Sussex, Texas A\&M University, and the OzDES Membership Consortium.

The DES data management system is supported by the National Science Foundation under Grant Numbers AST-1138766 and AST-1536171.
The DES participants from Spanish institutions are partially supported by MICINN under grants ESP2017-89838, PGC2018-094773, PGC2018-102021, SEV-2016-0588, SEV-2016-0597, and MDM-2015-0509, some of which include ERDF funds from the European Union. IFAE is partially funded by the CERCA program of the Generalitat de Catalunya.
Research leading to these results has received funding from the European Research
Council under the European Union's Seventh Framework Program (FP7/2007-2013) including ERC grant agreements 240672, 291329, and 306478.
We  acknowledge support from the Brazilian Instituto Nacional de Ci\^encia
e Tecnologia (INCT) do e-Universo (CNPq grant 465376/2014-2).

%(NOIRLab Prop. ID 2012B-0001; PI: J. Frieman)
Based in part on observations at Cerro Tololo Inter-American Observatory at NSF's NOIRLab, which is managed by the Association of Universities for Research in Astronomy (AURA) under a cooperative agreement with the National Science Foundation.

This work has made use of data from the European Space Agency (ESA) mission {\it Gaia} (\url{https://www.cosmos.esa.int/gaia}), processed by the {\it Gaia} Data Processing and Analysis Consortium (DPAC, \url{https://www.cosmos.esa.int/web/gaia/dpac/consortium}).
Funding for the DPAC has been provided by national institutions, in particular the institutions participating in the {\it Gaia} Multilateral Agreement.

This paper is based on data collected at the Subaru Telescope and retrieved from the HSC data archive system, which is operated by the Subaru Telescope and Astronomy Data Center (ADC) at NAOJ. Data analysis was in part carried out with the cooperation of Center for Computational Astrophysics (CfCA), NAOJ. We are honored and grateful for the opportunity of observing the Universe from Maunakea, which has the cultural, historical and natural significance in Hawaii. 

This research uses services or data provided by the Astro Data Lab, which is part of the Community Science and Data Center (CSDC) Program of NSF NOIRLab. NOIRLab is operated by the Association of Universities for Research in Astronomy (AURA), Inc. under a cooperative agreement with the U.S. National Science Foundation.

This manuscript has been authored by Fermi Forward Discovery Group, LLC under Contract No.\ 89243024CSC000002 with the U.S. Department of Energy, Office of Science, Office of High Energy Physics.

All analysis in this work was enabled greatly by the following software: \textsc{Pandas} \citep{Mckinney2011pandas}, \textsc{NumPy} \citep{vanderWalt2011Numpy}, \textsc{SciPy} \citep{Virtanen2020Scipy}, \textsc{Matplotlib} \citep{Hunter2007Matplotlib}, and \textsc{GetDist} \citep{Lewis:2019:Getdist}. We have also used
the Astrophysics Data Service (\href{https://ui.adsabs.harvard.edu/}{ADS}) and \href{https://arxiv.org/}{\texttt{arXiv}} preprint repository extensively during this project and the writing of the paper.

\section*{Data Availability}

All catalogs and derived data products (data vectors, redshift distributions, calibrations etc.) for the cosmology analysis are now publicly available through the Noirlab Datalab portal \citep{Fitzpatrick:2014:DataLab, Nikutta:2020:DataLab} as well as through Globus and other avenues. Please visit \url{dhayaaanbajagane.github.io/data_release/decade} for a list of the available dataproducts and their corresponding data access. Our intention is to make all useful products immediately available to the community. Please reach out to DA if a data product of interest to you is not on the above list.

\bibliographystyle{mnras}
\bibliography{References}

%%%%%%%%%%%%%%%%%%%%%%%%%%%%%%%%%%%%%%%%%%%%%%%%%%

%%%%%%%%%%%%%%%%% APPENDICES %%%%%%%%%%%%%%%%%%%%%

\appendix 

\section{Reanalyzing DES Y3 with the \decade pipeline}\label{appx:DES_matched}

\begin{figure}
    \centering
    \includegraphics[width = 0.9\columnwidth]{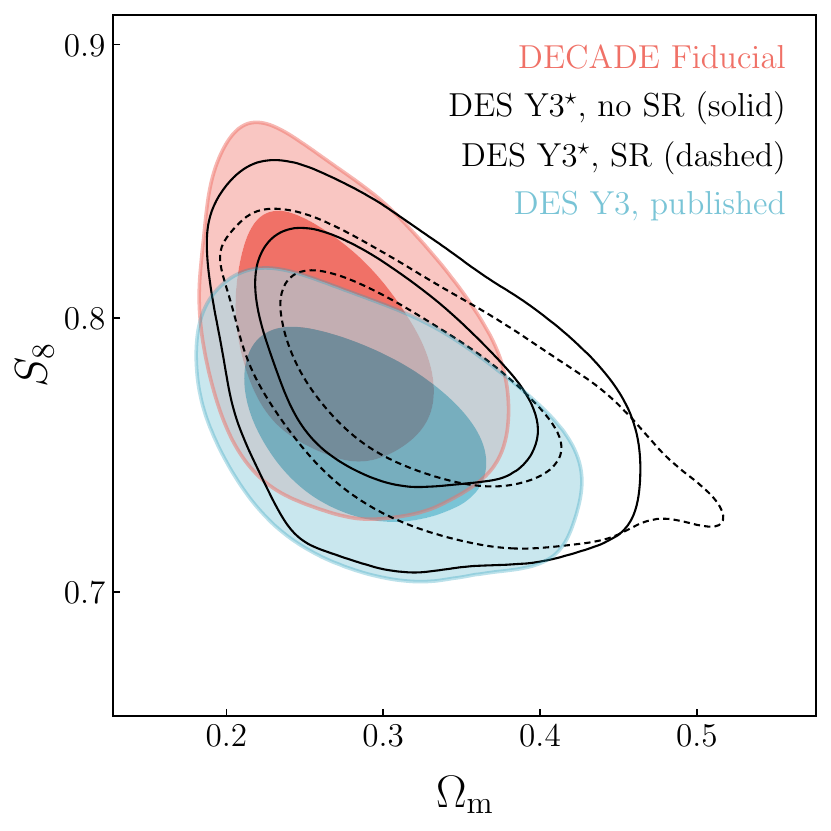}
    \caption{Constraints on $\Om$ and $\Seight$ from \decade and DES Y3. We show three variants of DES Y3. The DES constraints with $\star$ denote results from DES Y3 reanalysed using our inference pipeline. SR stands for the shear ratio measurements \citep{Sanchez2022} and is not available in the \decade data. The constraints from our reanalysis of DES Y3 differ from the published constraints due to a few reasons discussed in Section \ref{appx:DES_matched}, and the contour sizes differ as our analysis pipeline does not include SRs. See text for more details.}
    \label{fig:Y3_matched}
\end{figure}

When comparing against --- and combining with --- the constraints from DES Y3 cosmic shear, we follow the philosophy of \citet{DESKiDS2023} and reanalyze the DES Y3 data vector with the same inference pipeline used for the \decade data. The main difference between our pipeline and that of DES Y3 are as follows, in decreasing order of importance: (i) we do not use any shear ratio information \citep[SR,][]{Sanchez2022} as this is not available in \decade, (ii) we follow \citet{DESKiDS2023} in using the nonlinear matter power spectrum model of \citet{Mead2020a, Mead2021b} rather than that of \citet{Takahashi2012}, and; (iii) we use a slightly narrower prior for the different IA parameters, using $X \in [-4, 4]$ instead of $[-5, 5]$. The difference from (ii) shifts the constraint on $S_8$ slightly (see Table \ref{tab:constraints}), as documented in \citet{Secco2021} and \citet{DESKiDS2023}, while that from (i) widens the posterior width by $1.8 \times$, relative to the published result of DES Y3 \citep{Secco2021, Amon2021}. The latter is expected as SRs are known to help constrain IA parameters and self-calibrate redshift uncertainties, thereby improving the constraints on cosmic shear cosmology \citep[\eg][see their Figure 15 and Appendix B]{Amon2021}. The effect is significant when using the TATT model as this model has 5 free parameters, but is still prominent when using NLA as well. Figure \ref{fig:Y3_matched} shows the contours for the different variants discussed above. We have also verified (result not shown) that we reproduce the published constraints from DES Y3 if we modify our pipeline to match their analysis choices.

\section{Additional parameter constraints}\label{appx:more_param}

\subsection{Intrinsic Alignments}\label{appx:IA}

\begin{figure*}
    \centering
    \includegraphics[width = 1\columnwidth]{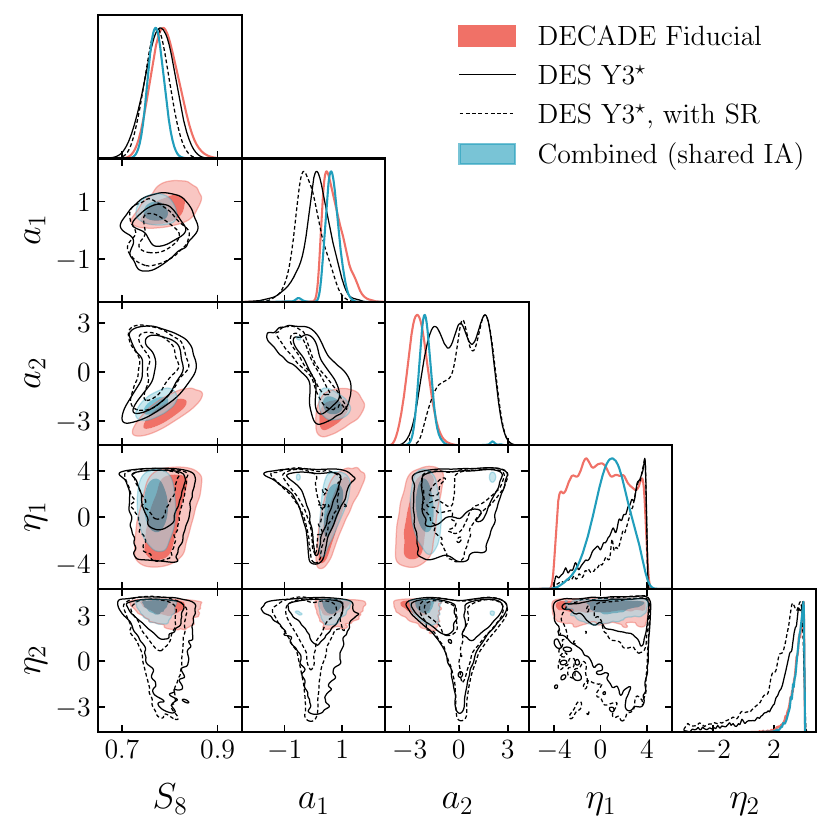}
    \includegraphics[width = 1\columnwidth]{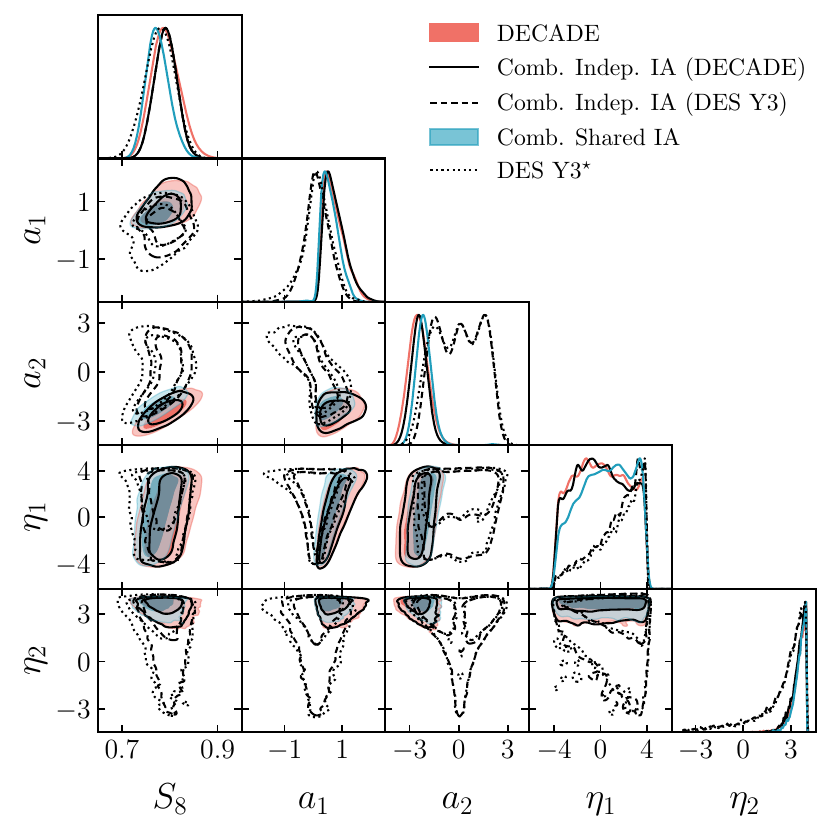}\\[10pt]
    \includegraphics[width = 1\columnwidth]{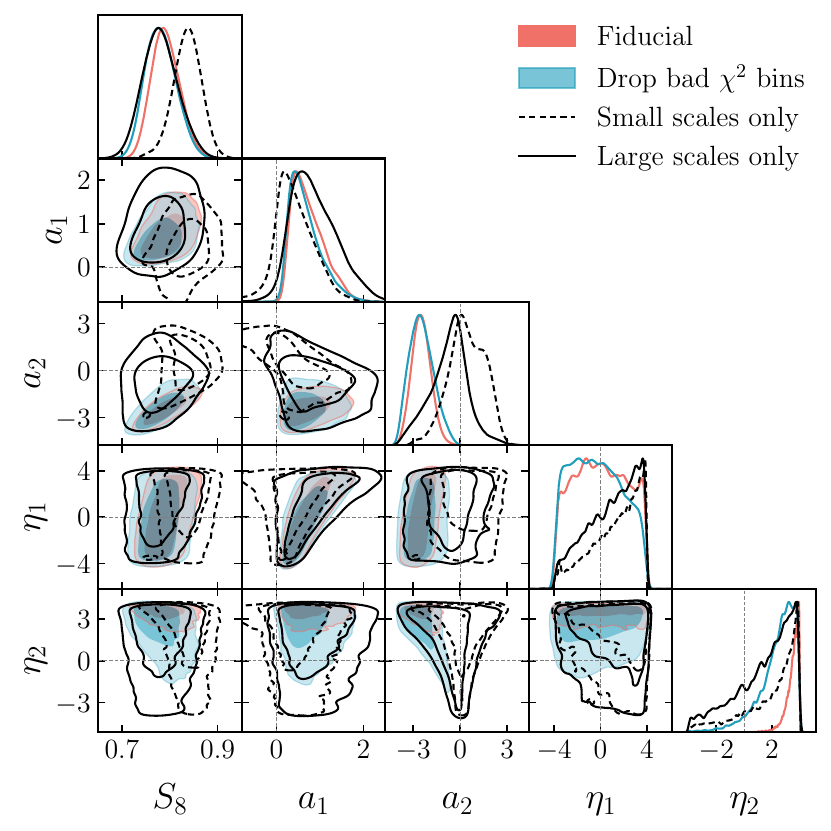}
    \includegraphics[width = 1\columnwidth]{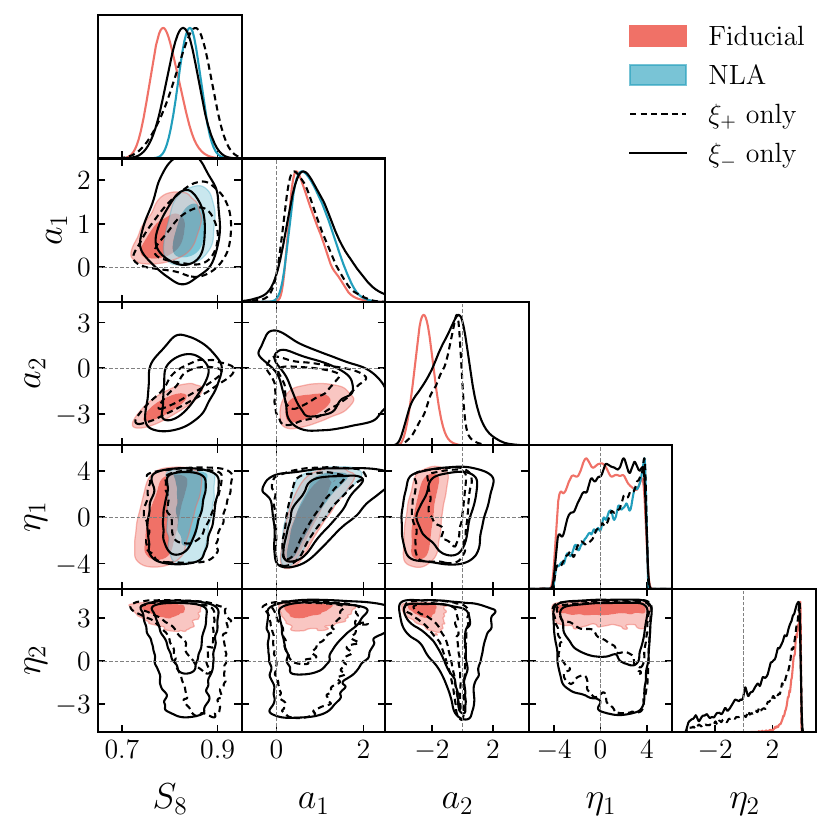}
    \caption{Constraints on the IA parameters for different analysis setups and different datasets, all under the \LCDM model. \textbf{Top Left:} the IA constraints for the results shown in Figure \ref{fig:constraints}. \textbf{Top right:} IA constraints for different subsets of the \decade and DES Y3 combined analysis. The black solid/dashed lines show the results from the ``independent IA'' setup where each dataset has its own TATT parameters, while the blue contours show the results from using a common set of parameters for both. \textbf{Bottom}: Constraints for different variants of the \decade analysis. The constraints on $a_1$ are similar between TATT and NLA --- and are also fairly similar if we exclude any subset of our data --- but those on $a_2$ broaden significantly and become consistent with $a_2 = 0$ once we exclude some subsets of the data. Notably, excluding the three bin pairs that contribute most to increasing our $\chi^2$ (see Section \ref{sec:sec:gof}) have IA posteriors that are very similar to those from our Fiducial analysis.}
    \label{fig:IA}
\end{figure*}

\begin{figure}
    \centering
    \includegraphics[width = 1\columnwidth]{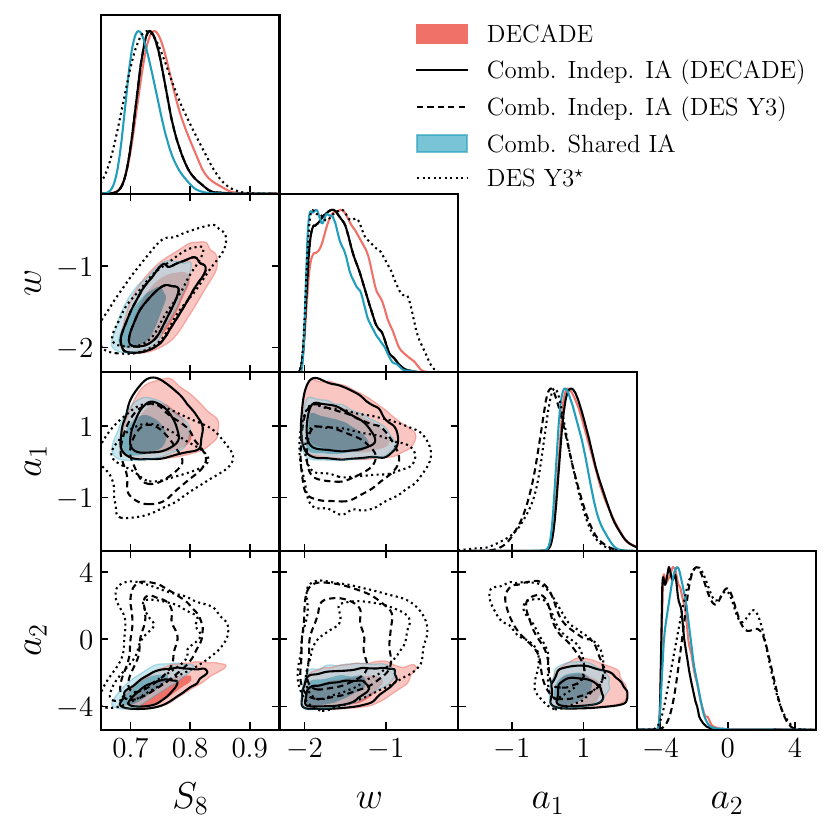}
    \caption{Similar to the top right panel of Figure \ref{fig:IA} but for \wCDM. There is a mild correlation between $a_2$ and  $w$ for the \decade IA parameters. We show two cases --- ``Independent IA'' where \decade and DES have their own set of TATT parameters (black solid, and black dashed, respectively), and ``Shared IA'' where they share a common set of five TATT parameters (blue).}
    \label{fig:IA_wcdm}
\end{figure}

The top left triangle plot of Figure \ref{fig:IA} show constraints on the IA parameters from \decade and DES Y3. As discussed previously, the \decade constrains prefer non-zero values for the IA parameters. These values are consistent with the (broad) posteriors from DES Y3 both when using/discarding the shear ratio measurements (SRs). Though in the former case, there is degraded consistency for the value of $a_2$.

The top right panels of the same figure show the behavior of the IA parameters in our combined analysis, including a version where the two datasets share a common set of TATT parameters. We show the \decade-only and DES Y3-only results for comparison. The combined analysis, when the IA model is not shared, finds similar posteriors to those of the individual surveys. The ``Shared IA'' result, where the IA model \textit{is} shared, has the same IA posteriors as the \decade-only result. This is expected as \decade has a strong preference for non-zero values in the TATT amplitudes $a_1, a_2$, with posteriors that are still broadly consistent with those of DES Y3. Thus, the combined analysis moves to the region of parameter space preferred by \decade since DES Y3 can still accommodate such values. Interestingly, the $p$-value of the DES Y3 best fit changes negligibly (from $p = 0.22$ to $p = 0.23$) between the analyses with independent/shared IA parameters. The same for \decade changes from $p = 0.02$ to $p = 0.015$. The best fit for the combined data vector (with scale cuts) corresponds to $p = 0.03$. The final constraints from the ``Shared IA'' model are,
\begin{align}
    \Seight &= 0.775^{+0.021}_{-0.026},\\
    \Om &= 0.298^{+0.038}_{-0.049},
\end{align}
which is consistent with the other lensing constraints discussed in Section \ref{sec:Constraints}.

Next, the bottom panels of Figure \ref{fig:IA} show the IA contours for a few different variant analyses (see Figure \ref{fig:data_variation}). First and foremost, the IA constraints are unchanged even if we exclude the three bin pairs contributing to our poor $\chi^2$ (see Section \ref{sec:sec:gof}). Thus, those bin pairs do not drive any specific preference in the IA parameter space. We also show results from excluding different subsets of the data. In all cases, the constraints on $a_1$ are fairly similar, but those on $a_2$ change noticeably. In particular, excluding a subset results in a broader posterior that is consistent with $a_2 = 0$. This highlights that the preference for $a_2 \neq 0$ comes from a combination of different parts of the data vector. We explore this more in Figure \ref{fig:IA_contrib}.

Figure \ref{fig:IA_wcdm} shows the IA amplitudes $a_1, a_2$ for the \wCDM analysis. There is a somewhat weak correlation between $a_2$ and $w$, and the negative values of the former, as preferred by the data, can slightly push $w$ to more negative values as well. Figure \ref{fig:IA_wcdm} also shows the constraints from using a shared set of IA parameters for the combined analysis, which are
\begin{align}
    \Seight &= 0.725^{+0.019}_{-0.033}\\[4pt]
    \Om &= 0.257^{+0.035}_{-0.052}\\[4pt]
    w &= -1.61^{+0.37}_{-0.13}.
\end{align}
This constraint on $w$ is consistent with \LCDM within $1.6\sigma$. The combined constraint is slightly improved, as expected, when using a common set of IA parameters for both datasets.

\begin{figure}
    \centering
    \includegraphics[width = 1\columnwidth]{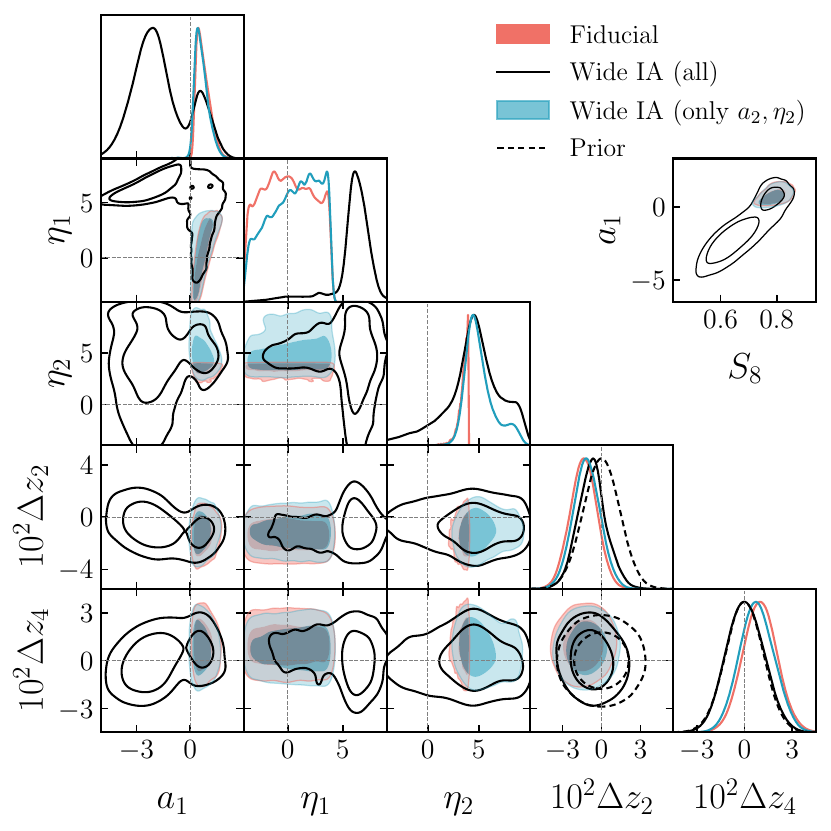}
    \caption{Similar to the top right panel of Figure \ref{fig:IA} for different choices of priors on the IA parameters (under the \LCDM model). Setting wide priors of $-10 < x < 10$ on $a_2, \eta_2$ alone (blue contour) gives constraints on $S_8$ that are consistent with the Fiducial analysis. If we set wide priors on all IA parameters (black solid), the model overfits to force the posteriors of $\Delta z_i$, the redshift calibration parameters, to agree with their priors (black dashed line). We consider this to be overfitting as the $\Delta z_i$ posteriors for the Fiducial setup are already consistent with the prior within $0.8\sigma$. The $a_1 - S_8$ plane in the top right shows how the IA bimodality propagates into $S_8$. The priors are shown only for $\Delta z_i$.}
    \label{fig:IA_wide}
\end{figure}

Next, we address the fact that the posteriors on $\eta_1$ and $\eta_2$ are limited by their prior ranges; note that the same behavior is also found in the DES Y3 cosmic shear analysis \citep[][see their Figure 17]{Secco2021}. We run a \decade variant analysis (``Wide-IA'') where the prior is significantly wider, $-10 < x_{\rm IA} < 10$ for all IA parameters $x$. In this case, the constraints occupy extreme values, introduce bimodality into the posteriors --- and into $S_8$ as a result --- but \textit{do not improve the goodness-of-fit compared to the Fiducial analysis}.

We have identified the cause of the bimodality is solely in $a_1$ and $\eta_1$, as the flexibility in those parameters is being used to force the posteriors of the redshift calibration parameters $\Delta z_i$, to match their prior (see Figure \ref{fig:IA_wide}). This matching boosts the posterior probability in the extreme regions of IA parameter space and causes the bimodal constraints. However, the posteriors of $\Delta z_i$ in our Fiducial constraint are still \textit{consistent with the priors} --- the mean value of the posterior is at-most $<0.8\sigma$ from $\Delta z_i = 0$ --- so the extreme values in our Wide-IA analysis are not compensating for any significant bias found in the redshift calibration. We therefore interpret the ``Wide-IA'' result as an ``overfitting'' scenario. Finally, we confirm that widening the priors on only $a_2$ and $\eta_2$ causes no bimodality and gives a constraint of $S_8 = 0.786^{+0.027}_{-0.031}$. This mean value \textit{and posterior width} are completely consistent with our Fiducial result. For this reason, we are comfortable using our current IA priors for all cosmological inference.

\begin{figure*}
    \centering
    \includegraphics[width = 2\columnwidth]{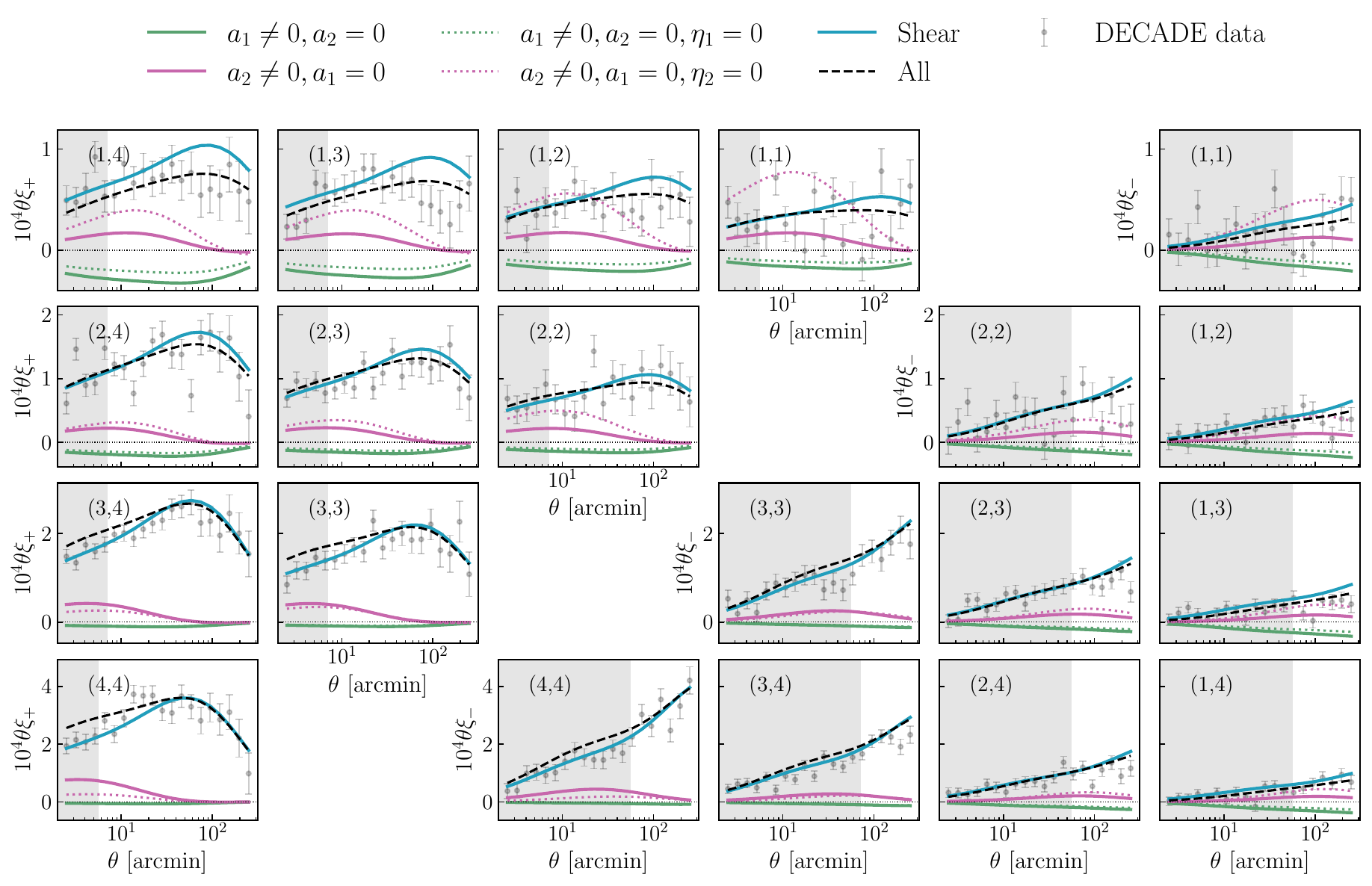}
    \caption{The contributions of different IA parameters to the final $\xi_\pm$ best-fit prediction under \LCDM. We show the IA contributions from nulling out specific amplitudes, and from nulling out only the redshift scaling of a given amplitude (see legend). The fiducial prediction is given as a black, dashed line alongside the \decade data (gray points). The prediction with no IA contribution is shown as the blue line. For the $\xi_+$ data vector, the $a_1$ contribution suppresses power on large scales and at low redshift, whereas the $a_2$ contribution amplifies power on small scales at high redshift. The $a_1, a_2$ terms cancel their contributions somewhat for the small scales at lower redshifts. The preference for $\eta_2 > 0$ suppresses the quadratic, $a_2$ IA terms at low redshift (purple dotted vs. solid), whereas $\eta_1$ has negligible impact in comparison. The $\xi_-$ data vector finds similar behavior to $\xi_+$ but at a somewhat suppressed level. Most notably, it also prefers a slight suppression of power on large scales. The $a_2$ contributions are mostly negligible to $\xi_-$ as the former's impact is on angular scales already discarded by our scale cuts (gray band).}
    \label{fig:IA_contrib}
\end{figure*}

Under the fiducial priors, the IA posteriors prefer $a_1 > 0$ and  $a_2 < 0, \eta_2 > 0$. In Figure \ref{fig:IA_contrib}, we explore the origin of these preferences. We compute best-fit predictions, but now fixing a subset of the IA amplitudes to zero.\footnote{Note that $a_1 = 0$ implicitly means $b_{\rm TA} = 0$ as the latter parameter rescales the former for one of the TATT contributions \citep[\eg][see their Equation 26]{Secco2021}.} The rest of the parameters are assigned their values from our Fiducial best-fit constraint. This helps isolate different behaviors in the five-parameter IA model. We first focus on distinguishing between the ``linear'' terms (green, dependent on $a_1$) and the ``quadratic'' terms (purple, dependent on $a_2$).\footnote{We use the terms ``linear'' and ``quadratic'' only heuristically. Our nomenclature is not strictly correct, in particular because $a_1$ still controls the amplitude of some terms that are quadratic in the density field; see Equations 21-23 in \citet{Secco2021}.}

The value $a_1 > 0$ is preferred as the linear terms (i) reduce power on larger scales for Bins 1 and 2, primarily in $\xi_+$ but somewhat in $\xi_-$ as well, and; (ii) reduce power on all scales in bin pairs (1,4) and (1,3). The auto/cross-correlations with Bin 1 and Bin 2 see the largest effect, which are also the measurements with an inherently lower signal-to-noise. The quadratic term has effectively no contribution on these larger angular scales; particularly in $\xi_+$, where it is essentially zero. This is consistent with Figure \ref{fig:IA} where the large scales-only analysis prefers $a_2 = 0$. Instead, the main impact from the quadratic contribution is on small scales, and across all bins. We are unable to isolate which combination of data points induce this overall preference. Note that this term also amplifies $\xi_-$ on smaller scales, but such scales are already removed by our scale cuts (gray bands) and have no impact on our final constraints. This shows that the preference for $a_2 \neq 0$ is predominantly from the higher-signal-to-noise $\xi_+$ measurement rather than the $\xi_-$ one, as is also found in Figure \ref{fig:IA}. To summarize, the IA model finds $a_1 > 0, a_2 < 0$ to accommodate a mild scale dependence preferred by the data.\footnote{ We have checked, using simulated data vectors, that catastrophic redshift errors --- for example, if our mean redshift is wrong by $\Delta z = 0.1$, which is nearly ten times worse than our calibrated priors --- do not generate a scale dependence that mimics the one predicted by the best-fit IA model.}

Having identified the need for non-zero amplitudes in the linear/quadratic IA terms, we then check the redshift dependence associated with these amplitudes. We recompute IA contributions but fix the redshift power-law indices, $\eta_1, \eta_2 = 0$. These results are shown as dotted lines in Figure \ref{fig:IA_contrib}. The linear term's redshift scaling, $\eta_1$, has very little impact given that scaling parameter is mostly unconstrained (Figure \ref{fig:IA}). However, the same scaling for the quadratic term is found to prefer $\eta_2 > 0$ as mentioned above. Comparing the dashed and solid purple lines in Figure \ref{fig:IA_contrib} shows that the low-redshift datapoints motivate this preference. Specifically, they prefer a smaller amplitude for the quadratic contribution, and so $\eta_2$ must take large values to suppress the relevant amplitude at such redshifts (see Equation \ref{eqn:IA_ampl}). While Figure \ref{fig:IA_contrib} shows that different subsets of the data vector are driving the preference in different IA parameters, we show that reanalyzing our data using/discarding such subsets results in $S_8$ constraints that are still within $1\sigma$ of our Fiducial result (Section \ref{sec:sec:dropdata}).

In summary, we have undertaken numerous, extensive tests of the IA posteriors to understand the origin of the parameter constraints and check the possibility of systematics driving these constraints. Our tests, the primary of which are the reanalyses of the forty-six catalog-level splits \citepalias{paper3}, find consistent results for the IA constraints and do not isolate any particular systematic as a potential cause for the observed behavior. It is therefore still possible that our constraints reflect (at least partly) a real IA signal in the data. It also remains possible that some part of the constraint is driven by a non-IA contribution that is not probed by our forty-six splits. Regardless, we stress that the $S_8$ constraints from our numerous IA-related tests are \textit{consistent with our Fiducial results} and highlight the robustness of our quoted constraints. Subsequently, the extensive discussion of IA necessitated in this work further highlights the subtleties in constraining and \textit{understanding} the IA signal in a lensing sample without the use of additional data, \eg spectroscopic datasets like those used in \citet{Samuroff2022} or photometric ``lens'' samples like those used in \citet{Samuroff:2019:IA} and \citet{Samuroff2022}.

\subsection{Small scales and baryons}\label{appx:Tagn}

\begin{figure}
    \centering
    \includegraphics[width = 1\columnwidth]{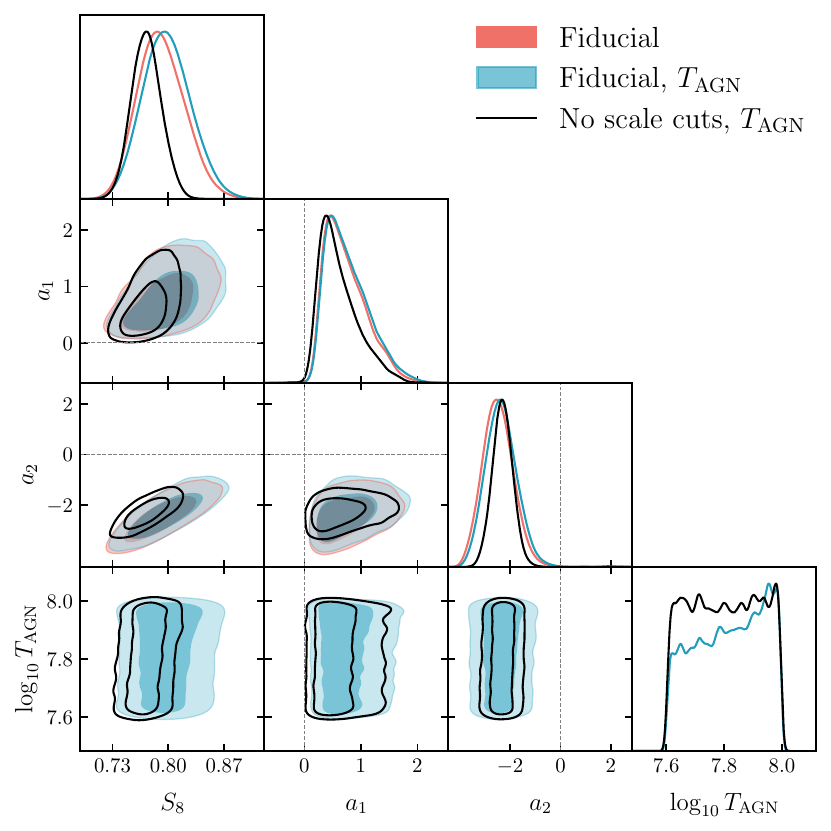}
    \caption{\LCDM Constraints from \decade after including baryon corrections in the matter power-spectrum using the model of \citet{Mead2021b}, and after removing all scale-cuts in addition to the baryon modeling. The parameter $T_{\rm AGN}$ is an effective feedback strength, and higher values correspond to stronger suppression of the power spectrum; see \citet{Mead2021b} for more details. The constraints from all analysis choices are consistent, and there is no coupling between the IA and $T_{\rm AGN}$. The gray dotted lines denote the ``no IA'' case of $a_1, a_2 = 0$.}
    \label{fig:Tagn}
\end{figure}

Our scale cuts are primarily motivated by the uncertainty in the nonlinear power spectrum model on small scales. 
The evolution of structure on these small scales is significantly altered by the non-gravitational processes of baryons; predominantly, this is the ejection of gas outside of the halo due to energetic feedback \citep{Chisari2018BaryonsPk}. Hydrodynamical simulations \citep[see][for a review]{Vogelsberger2020Hydro}, which model these effects through approximate subgrid prescriptions, generate a variety of predictions for the properties of halos in simulations \citep[\eg][]{Anbajagane2020StellarProp, Lim2021GasProp, Lee2022rSZ, Stiskalek2022TNGHorizon, Anbajagane2022Baryons, Anbajagane2022GalaxyVelBias, Shao2022Baryons, Shao2023Baryons, Gebhardt2023CamelsAGNSN} which translates into a variety of predictions for changes in the nonlinear matter power spectrum \citep[\eg][]{vanDaalen2011, Chisari2018BaryonsPk, Amon2022S8Baryons}. Such variations can be represented through phenomenological, halo-based models such as ``baryonification'' \citep{Schneider:2015:Baryons, Schneider2019Baryonification, Arico:2021:Bacco, Anbajagane:2024:Baryonification} and also a variant of the \textsc{HMCode2020} model \citep{Mead2021b}. The latter includes a phenomenological parameter, $T_{\rm AGN}$, that quantifies the baryonic contribution to the nonlinear matter power spectrum, and this parameter was varied in the joint-analysis of DES and KiDS \citep{DESKiDS2023}.

We mimic the setup of the joint-analysis of DES and KiDS, and run two analysis variants where we explicitly model baryon corrections through the $T_{\rm AGN}$ parameter mentioned above. In particular, we run one variant where we use the same scale cuts as the Fiducial analysis (blue) and another where we use all available scales in the data vector (black solid). In both cases, we use the prior $T_{\rm AGN} \in [10^{7.6}, 10^{8.0}]$ which is the range the model is calibrated for. The posterior is shown in Figure \ref{fig:Tagn}. The $T_{\rm AGN}$ is unconstrained, similar to findings from \citet{DESKiDS2023, Bigwood:2024:BaryonsWLkSZ}. Allowing freedom in the $T_{\rm AGN}$ parameter does not change any of the IA amplitudes, and shifts $S_8$ (relative to the Fiducial constraints) by less than $\approx\!0.5\sigma$. The weak coupling of IA and baryons has also been previously found in simulation measurements \citep[\eg][]{Tenneti:2017:BaryonIA, Soussana:2020:BaryonIA}.

\subsection{Full parameter space}\label{appx:fullparam}

\begin{figure*}
    \centering
    \includegraphics[width = 1.99\columnwidth]{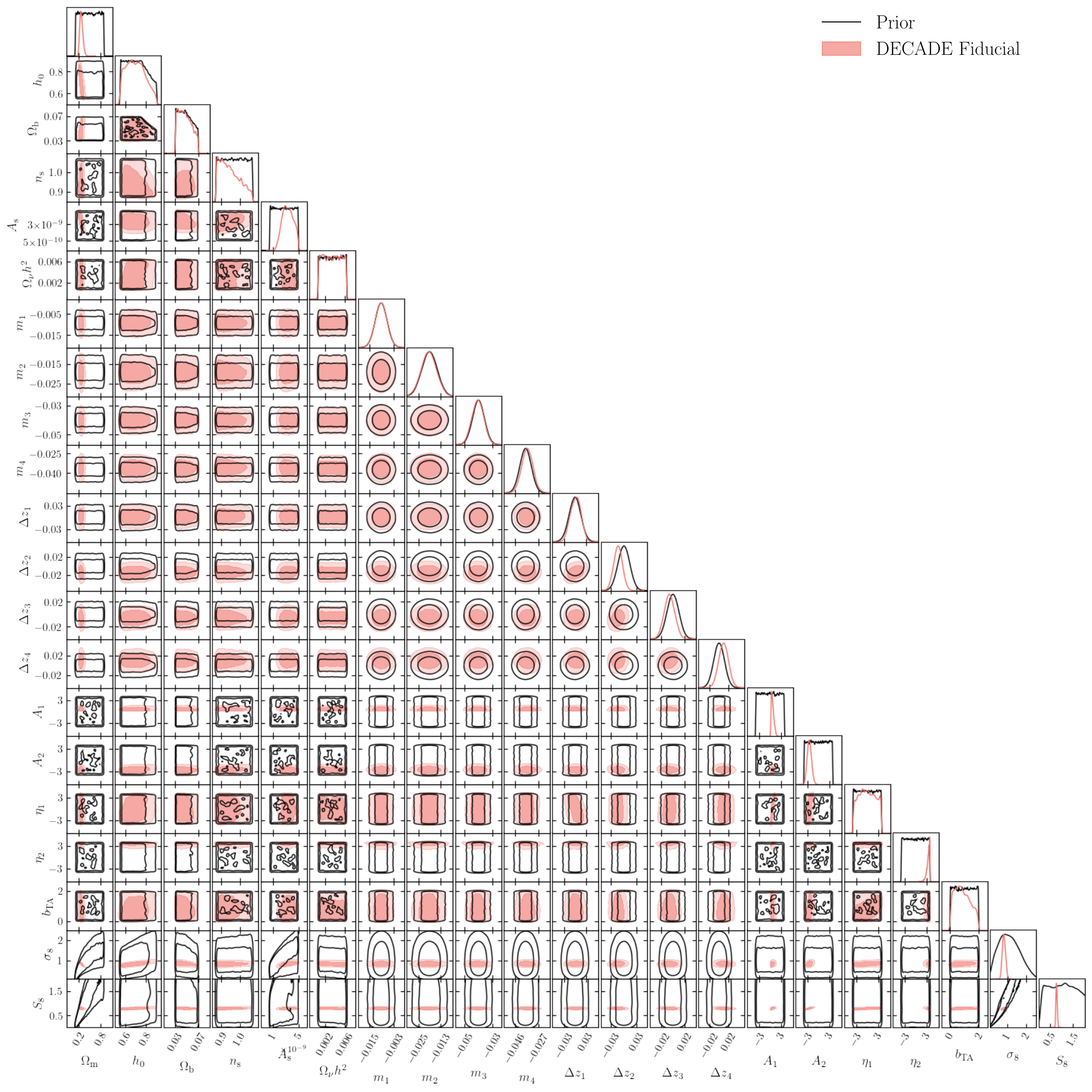}
    \caption{Full parameter constraint for the fiducial analysis (Section~\ref{sec:lcdm}, Figure~\ref{fig:constraints}). We also plot the priors associated with this chain. }
    \label{fig:full_param}
\end{figure*}

In Figure~\ref{fig:full_param} we show posteriors from the full parameter space in our fiducial $\Lambda$CDM analysis (red) and the prior that we sample from (black). We discuss the behavior of the intrinsic alignments (IA) parameters extensively in Section~\ref{sec:sec:IA} and Appendix~\ref{appx:IA}, so we do not elaborate further in this section.

\section{Unblinding tests and findings}
\label{appx:unblind_test}

For this work, we make all the analysis choices (the model, priors, and scale cuts) using simulated data vectors. These relevant tests/analyses are found in \citetalias{paper3}. Once the analysis choices are fixed, we perform a number of checks on the blinded data vectors (see Section \ref{sec:unblinding} for details on our unblinding method). These tests, together with all the previous tests on the data and the model, cross-check the robustness of our results and justify unblinding the data vector. The full criteria for unblinding is summarized below: 

\begin{itemize}
    \item \textbf{Shear catalog tests:} All tests described in \citetalias{paper1} must pass. These include a number of tests that show the galaxy shape catalog does not contain significant contamination from point-spread function modeling, and does not correlate with image data quality. We also test for $B$-modes and tangential shear around locations where we do not expect cosmological signals.
    \item \textbf{Redshift tests:} All tests described in \citetalias{paper2} must pass.  These primarily include a check that the redshift distribution estimated through the SOMPZ method is consistent with that estimated from clustering redshifts.
    \item \textbf{Methodology tests:} All tests described in \citetalias{paper3} must pass. This includes checking that the cosmological inference pipeline is robust to different analysis choices. We also perform forty-six different end-to-end tests using subsets of the data split by observing conditions or galaxy properties, and check that the resulting cosmological constraints are consistent with our fiducial results. Our criteria was $3\sigma$ consistency, but we find all tests pass within $2\sigma$ with most passing within $1\sigma$.
    \item \textbf{Final tests (with blinded $\Lambda$CDM chains):} 
    \begin{itemize}[label=$\circ$]
        \item We visually inspect the constraints in the full parameter space to see that posteriors on the relevant parameters (\eg $\Omega_{\rm m}$, $S_8$, $a_1$, $a_2$) are not pushing against the priors in an unexpected way.\footnote{Some IA parameters, such as $\eta_1$ and $\eta_2$ are expected to be prior-dominated \citep{Secco2021, Amon2021} and were not used in this test. See Appendix \ref{appx:IA} for discussions on the chosen prior ranges for these parameters.} We also checked that our nuisance parameters, $m_i$ and $\Delta z_i$, had posteriors that were consistent with the priors, \ie the final constraints did not shift to extreme values in the prior. 
        \item The goodness-of-fit for our best-fit model has a $p$-value $>0.0015$ which corresponds to the model and data being consistent within $<3\sigma$.
        \item When comparing constraints on $S_8$ from the Fiducial analysis, to those obtained by excluding individual redshift bins, the differences in the constraints are less than $3 \sigma$.\footnote{We realized post-unblinding that 3$\sigma$ is likely too inclusive a threshold for this test given the contours will be fairly correlated. However, we note that in practice all the contours result in shifts less than $\sim 1\sigma$, so our choice of a more conservative threshold did not have an impact on the result of this test. See Figure \ref{fig:data_variation}.}
    \end{itemize}
\end{itemize}

We define ``parameter shifts/differences'' using the simple distance metric in the $S_8$ direction, assuming Gaussian posteriors: 
\begin{equation}\label{eq:s8_dist}
    \text{number of }\sigma = \frac{[S_8]_1 - [S_8]_2}{\sqrt{\sigma([S_8]_1)^2+\sigma([S_8]_2)^2}},    
\end{equation}
where $[S_8]_i$ and $\sigma([S_8]_i)$ are the mean and standard deviation, respectively, of the posterior for constraint $i$, and the subscripts, 1 and 2, refer to the two posteriors from which we obtain the value of interest. We note that more complex and rigorous distance/tension metrics exist \citep[\eg][]{Doux:2021:tensions, Raveri:2021:NGTension}, but since most of the constraining power in cosmic shear lies in the $S_8$ parameter, and the $S_8$ constraints are fairly Gaussian, this simple metric is sufficient to capture most of the relevant information (\ie deciding whether a change in $S_8$ is significant). 

The majority of tests above passed fairly trivially. However, there were two results that still pass our established criteria but differed from our prior expectations; for this reason we record them here. Before unblinding, we performed extensive tests associated with these individual results, and ensured we did not find any known source of lensing systematics that generated them:
\begin{itemize}
    \item The goodness-of-fit was slightly low (though still passing our criteria), and was lower when using the NLA IA model than the TATT model. We then decided to be conservative and use TATT as our fiducial IA model. We also note that using NLA or TATT resulted in $\approx\!\!1.3\sigma$ shift in cosmological parameters (discussed in Section~\ref{sec:Discussion}). Post-unblinding, we also used Bayesian evidence ratios to find our data showed a strong preference for the TATT model over NLA.
    \item We identified that our relatively high $\chi^2$ originates from the $\xi_+$ measurement in three bin pairs $(1,1)$, $(2,4)$, $(4,4)$. The origin of the high $\chi^2$ is not distinctly isolated in one redshift bin, and the residuals do not show any clear trend with scale. More importantly, we verified pre-unblinding that our cosmology constraints are consistent within $0.24\sigma$ when including/discarding these three bin pairs. Even after discarding these three bin pairs, switching our IA model from TATT to NLA shifts $S_8$ to higher values. Given all the above results/checks, we determined the relatively high $\chi^2$ could indeed be caused by statistical fluctuations and did not discard these specific bin pairs.
    \item Our constraints on IA found a significant, non-zero value of $a_2$, which is not unreasonable \citep[\eg][]{Samuroff:2019:IA} but is somewhat large relative to DES Y3 \citep{Secco2021, Amon2021}. We used simulated data vectors to find that the presence of noise alone can generate such spurious detection of IA; this is consistent with findings in DES Y3 \citet[][see their Figure 15 and Section B]{Amon2021}. We have since done extensive tests on the interaction of IA with other aspects of our analysis and discuss them extensively in Section \ref{sec:sec:IA} and Appendix \ref{appx:IA}.
\end{itemize}

After unblinding, we regenerated a covariance matrix using the best-fit cosmology from our Fiducial constraint. We then ran chains for our final cosmological constraints, and also reran a number of consistency tests to verify that the results did not change. During this process, we discovered one minor/negligible inconsistency in our data pipeline, one in our simulation tests pipeline, and one in our redshift test. We have since corrected them and all final results presented in this series of papers use the corrected versions. None of these changes affect the final constraints from the \decade dataset. The changes made post-unblinding are:
\begin{itemize}
    \item We discovered a minor inconsistency between the shear weights used in the calculation of $m$ from the image simulations \citepalias{paper1} and the shear weights used in all other applications (\eg computing effective number densities, data vector computations, \etc). This was corrected post-unblinding and the resulting change in $m$ shifted the constraint on $S_8$ by only $0.08\sigma$. All results presented below (and all tests mentioned above, where relevant) use the post-unblinding, corrected $m$ values.
    \item We discovered a mismatch in the cosmology used for the simulated data vector and the covariance matrix used in the simulation tests shown in \citetalias{paper3}. In addition, our simulated data vector was still using the NLA model and was not updated to our fiducial choice of TATT. We have since updated all the simulated data vectors and covariances, and have checked that all tests still pass after the update; all results shown in \citetalias{paper3} are from the post-correction simulated data vector.
    \item We improved our redshift test --- the $\chi^2$ between the SOMPZ and clustering redshift (WZ) estimates, shown in Figure 10 of \citetalias{paper2} --- by (i) incorporating uncertainties from the SOMPZ-based redshift samples into the final covariance, (ii) accounting for the free parameters of the SOMPZ-to-WZ forward model in defining $N_{\rm dof}$ for the $\chi^2$ test, and; (iii) limiting the redshift range of the test to $0 < z < 1.6$, as the SOMPZ $n(z)$ has minimal/no support beyond $z > 1.6$. The updated test passes with greater probability than was the case for the pre-unblinding iteration, and the updates above do not affect the fiducial SOMPZ distributions or the alternative ``SOMPZ+WZ'' distributions used in this work. See Figure \ref{fig:nofz}, for these $n(z)$ estimates.
\end{itemize}

% Don't change these lines
% \bsp	% typesetting comment
\label{lastpage}
\end{document}